\documentclass[aps,nofootinbib,amsmath,prd,onecolumn,notitlepage,showpacs,superscriptaddress,11pt]{revtex4-1}

\usepackage[american]{babel}
\usepackage{amsfonts}
\usepackage{amsmath}
\usepackage{amssymb}
\usepackage{slashed}
\usepackage{upgreek}

\usepackage{xcolor}
\usepackage{bm}
\usepackage{float}
\usepackage[utf8]{inputenc}
\usepackage{nicefrac}
\usepackage{hyperref}
\usepackage{url}
\usepackage{tikz-cd}
\usepackage[normalem]{ulem}
\usepackage{mathtools}
\usepackage{graphicx,txfonts}

\newcommand{\mycomment}[1]{}

\definecolor{darkblue}{rgb}{0.1,0.1,0.7}
\hypersetup{colorlinks,
           linkcolor={darkblue},
           citecolor={darkblue},
           urlcolor={darkblue}
}

\newcommand{\lvec}[2]{\raise #1\hbox{$^\leftarrow$} \hspace{-9pt} #2}
\newcommand{\rvec}[2]{\raise #1\hbox{$^\rightarrow$} \hspace{-9pt} #2}
\newcommand{\lrvec}[2]{\raise #1\hbox{$^\leftrightarrow$} \hspace{-9pt} #2}

\usepackage[normalem]{ulem}

\allowdisplaybreaks

\newcommand{\dd}{{\rm d}}

\DeclarePairedDelimiter\abs{\lvert}{\rvert}%
\DeclarePairedDelimiter\norm{\lVert}{\rVert}%

\makeatletter
\let\oldabs\abs
\def\abs{\@ifstar{\oldabs}{\oldabs*}}
\let\oldnorm\norm
\def\norm{\@ifstar{\oldnorm}{\oldnorm*}}
\makeatother

\makeatletter
\def\l@subsubsection#1#2{}
\makeatother

\makeatletter
\def\l@section#1#2{%
  \@dottedtocline{1}{1.5em}{2em}{#1}{#2}
  \addvspace{0.2em}} 
  \def\l@subsection#1#2{%
  \@dottedtocline{2}{2.7em}{2em}{#1}{#2}
  \addvspace{0.2em}}
\makeatother

\begin{document}

\title{Ambient space and integration of the trace anomaly}

\author{Gregorio Paci}
\email{gregorio.paci@phd.unipi.it}
\affiliation{
Dipartimento di Fisica and INFN - Sezione di Pisa, Universit\`a di Pisa, Largo Bruno Pontecorvo 3, 56127 Pisa, Italy}

\author{Omar Zanusso}
\email{omar.zanusso@unipi.it}
\affiliation{
Dipartimento di Fisica and INFN - Sezione di Pisa, Universit\`a di Pisa, Largo Bruno Pontecorvo 3, 56127 Pisa, Italy}

\begin{abstract}
%
We use the ambient space construction, in which spacetime is mapped into a special lightcone of a higher dimensional manifold, to derive the integrable terms of the trace anomaly in even dimensions. We argue that the natural topological anomaly is the so-called $Q$-curvature, which, when projected from the ambient space, always comes with a Weyl covariant operator that can naturally be adopted for the integration of the anomaly itself in the form of a nonlocal action.
The use of the ambient space makes trasparent the fact that there are some new ambiguities in the integration of the anomaly, which we now understand geometrically from the ambient point of view. These ambiguities, which manifest themselves as undetermined parameters in the integrated nonlocal action, become more severe in dimensions $d\geq 6$
and do not seem to be related to a choice of the renormalization scheme.
\end{abstract}

\maketitle

\vspace{-0.6cm}

\tableofcontents


\section{Introduction}
\label{sect:intro}

Conformal field theories (CFTs) in $d$-dimensional flat space enjoy an energy-momentum tensor $T_{\mu\nu}$, which is conserved and traceless, $T^\mu{}_\mu=0$. When a CFT is coupled to a metric $g_{\mu\nu}$ in curved space, which acts as a source for the (variational) energy-momentum tensor, conformal invariance is naturally generalized by Weyl invariance, assuming also invariance under reparametrizations in the form of diffeomorphisms. Classically, Weyl invariance implies that the trace of the energy momentum tensor is zero by virtue of the Noether identities, while the conformal group emerges as the subset of diffeomorphisms and Weyl transformations that leave the metric invariant, i.e., the conformal isometries.

In a path-integral approach, the quantization procedure relates the expectation value of the energy-momentum tensor $\langle T_{\mu\nu}\rangle$ to the variation of the effective action $\Gamma$ with respect to the metric, which is thus its source.
The procedure introduces an anomalous nonzero value to the expectation value of the trace of the energy-momentum tensor, $\langle T \rangle \equiv g^{\mu\nu} \langle T_{\mu\nu} \rangle\neq 0$, known as trace anomaly, assuming that diffeomorphisms invariance (and thus covariance) is preserved during quantization \cite{Duff:1993wm}.
The trace anomaly cannot be removed by changing the regularization scheme of the path-integral and, for this reason, it is a special property of the CFT that we have to live with. In fact, the coefficients of the anomaly manifest themselves through correlators that involve multiple copies of the energy-momentum tensors, as well as the other relevant operators of the theory, also in the flat space limit.

From the point of view of the curved space, the trace anomaly
can be parametrized in terms of sums and products of Riemannian curvatures. Given that the anomaly must satisfy specific consistency conditions, because it is the anomaly of an Abelian symmetry, it is thus possible to classify its terms according to their properties. The classification leads to the most important terms, known as Type-A and Type-B anomalies, which we refer to as $a$-anomalies and $b$-anomalies later in the paper \cite{Deser:1993yx}.

In this paper we show how to use the ambient space manifold by Fefferman and Graham \cite{Fefferman:2007rka}, which is a $(d+2)$-dimensional manifold inside which spacetime is projected onto a special lightcone known as the ``projective lightcone'', to construct and classify the aforementioned $a$- and $b$-anomalies. In fact, these turn out to be intimately connected
to both the intrinsic and extrinsic geometrical properties of the ambient manifold and the projective lightcone. The paper is structured in two main parts, because first we construct, almost pedagogically, all the necessary geometrical entities combining results from both the mathematics and physics literatures, and later we apply them to parametrization and integration of the anomaly in both general and specific cases. Our primary focus is the construction of the possible terms of the anomaly, while our secondary focus is the integration of such anomalies, and both applications are greatly simplified by the use of the ambient approach. The most pragmatic outcome of our analysis is that we give an ambient geometrical explanation to some ambiguities that emerge in the integration of the trace anomaly.

The paper is structured as follows. In Sect.~\ref{sect:ambient-main} we introduce the ambient manifold. For this purpose we have chosen a bottom-up approach in which we discuss the ambient manifold as the curved space generalization of the embedding space of CFTs. Our choice is motivated because it is relatively straightforward to show how scale invariance is manifested through the embedding formalism. We have also chosen to present several details in the construction and also tried to touch upon some points that to our knowledge are not discussed elsewhere, such as the extrinsic properties of the projective lightcone.

In Sect.~\ref{sect:integrable-terms} we use the ambient space as a tool to construct the necessary
$a$- and $b$-anomalies (our primary focus), which have to satisfy specific integrability conditions. The important point of our discussion is that the procedure that generates the $a$-anomaly produces at the same time a Weyl covariant differential operator, which is the necessary one to construct a Green function that can then be used in the integration of the anomaly. We find that this point, namely that the $a$-anomaly and its differential operator have a common origin, is not particularly stressed in the literature, but fundamental for the costruction of the effective action.

In Sects.~\ref{sect:anomaly-integration} and \ref{sect:ambiguities} we 
address our secondary focus of integrating the anomaly in the form of nonlocal actions, which is arguably the most important one for physical applications given that these actions have already been used in contexts \cite{Duff:1993wm}.
Specifically, in Sect.~\ref{sect:anomaly-integration} we discuss a general and natural approach to the integration of the trace anomaly, based on a modern view of the topic. Instead in Sect.~\ref{sect:ambiguities} we show how the procedure of integration of the anomaly may be ambiguous, for example in $d\geq6$, and give an ambient geometrical interpretation to such ambiguity.

In Sect.~\ref{sect:conclusions} we attempt a conclusion. In particular, we address one important shortcoming of the ambient space construction, as developed in this paper and in many other references, which involves how so-called Weyl ``obstructions'' are circumvented by means of what could be seen as an ``analytic continuation'' in the dimension $d$ to obtain the $a$- and $b$-anomalies. We try to address how such shortcoming may have affected the results, based also on the comparison with other approaches in the literature, and why it would be interesting to avoid them in future applications.

\subsubsection*{Notations and conventions}

\begin{itemize}
 \item $M$ is the spacetime with metric $g_{\mu\nu}$ and Levi-Civita connection $\nabla$. Greek indices are holonomic indices on $M$ for coordinates $x^\mu$, with the exception of $\rho$. Riemannian curvatures on $M$ are denoted with $R$.
 \item When a new ``metric'' $h_{\mu\nu}$ on $M$ is introduced with Levi-Civita connection $D$, its Riemannian curvatures are denoted with $R[h]$ to distinguish them from those of $g_{\mu\nu}$.
 \item $\tilde{M}$ is the ambient space with metric $\tilde{g}_{AB}$ and Levi-Civita connection $\tilde{\nabla}$. Capitalized Latin indices are holonomic indices on $\tilde{M}$ for coordinates $X^A$ (while $Y^A$ are reserved for lightcone Cartesian vector coordinates). Riemannian curvatures on $\tilde{M}$ are denoted with $\tilde{R}$.
 \item The symbols $\rho$ and $t$ are reserved for two special coordinates on $\tilde{M}$. When used as index labels they denote the correspoding coordinate dependent component. Partial derivatives with respect to $\rho$ are denoted with a prime, e.g., $\partial_\rho f \equiv \frac{\partial f}{\partial \rho}=f'$
 \item The conformal tensors ${\cal J}$, Schouten's $K_{\mu\nu}$, Cotton's $C_{\mu\nu\alpha}$, Weyl's $W_{\mu\nu\alpha\beta}$ and Bach's $B_{\mu\nu}$ are related to Riemannian curvatures in Appendix~\ref{sect:conformal-tensors}. Formulas of the first part of the paper are mostly given in terms of the conformal tensors, while formulas of the second part are mostly give in terms of the Riemannian tensors.
 \item If $\Delta$ is a differential operator, $\frac{1}{\Delta}$ is a shorthand for its Green function, $G_\Delta=\frac{1}{\Delta}$, which solves $\Delta_x G_\Delta(x,y) = \delta^{(d)}(x,y)$ for the $d$-dimensional covariant Dirac-delta $\int \dd^d x \sqrt{g} \, f(x)\, \delta^{(d)}(x,y)= f(y)$. Thus we have, for example
 $$
 \int \dd^d x \sqrt{g} \, R \frac{1}{\Delta} R = \iint\dd^d x \, \dd^d y \sqrt{g}_x \sqrt{g}_y \, R_x G_\Delta(x,y) R_y
 \,.
 $$
 \item ${\cal L}_\xi T$ is the Lie-derivative of the tensor $T$ in the ``direction'' of the vector $\xi$. If $\xi^\mu \partial_\mu = \partial_a$
 for some parameter $a$ that is also used as coordinate, then $({\cal L}_\xi T)_{\cdots} = \partial_a T_{\cdots}$ at the level of the components.
\end{itemize}

\section{The ambient space}
\label{sect:ambient-main}

In this section we give a motivated introduction of the ambient space formalism and outline many of its geometrical properties. As a starting point for the construction we have chosen the embedding formalism of CFTs \cite{Weinberg:2010fx,Costa:2011mg}, which is probably better-known in the theoretical physics community. A comprehensive review of the ambient space can be found in Ref.~\cite{Jia:2023gmk} which also generalizes it to include a gauge potential for Weyl symmetry.

\subsection{From the embedding space to the ambient space}\label{sect:cft}

The embedding space approach to CFT is based on the notion that the conformal group of $d$-dimensional flat space is isomorphic to the Lorentz group in $d+2$ dimensions. Its aim is then to construct an embedding of the $d$-dimensional CFT in $d+2$ dimensions such that Lorentz-covariance in the embedding implies conformal invariance in $d$ dimensions.

Consider a $d$-dimensional flat space $M$ (topologically $\mathbb{R}^d$) with metric $\eta=\eta_{\mu\nu} {\rm d}x^\mu {\rm d}x^\nu$
and a $(d+2)$-dimensional flat space $\tilde{M}$ with metric $\tilde{\eta}=\eta_{AB} {\rm d}X^A {\rm d}X^B$, both in Cartesian coordinates.
We present everything assuming the signature of $\eta$ to be the Minkowskian $(-,+,+,\cdots)$, but the construction is independent of the signature, as long as $\tilde{\eta}$ has one additional time direction in comparison to $\eta$.\footnote{%
In general, if $M$ has signature $(p,q)$ then $\tilde{M}$ has signature $(p+1,q+1)$.
}
Inside $\tilde{M}$, we identify a projective lightcone, denoted $\Sigma$, through the conditions
\begin{align}\label{eq:projective_lightcone}
 \eta_{AB} Y^A Y^B= 0 \, , \qquad { \rm and} \qquad Y^A \sim \lambda Y^A \qquad {\rm for} \qquad \lambda \in \mathbb{R}
 \, .
\end{align}
The projective lightcone is a codimension-$2$ null hypersurface of $\tilde{M}$, that is, it has dimension $d$ rather than $d+1$, which is what one would expect for timelike or spacelike hypersurfaces. This is thanks to the equivalence relation $ Y^A \sim \lambda Y^A$, itself caused by the fact that $Y^A$ are null.
The aim now is to construct a one-to-one correspondence between $M$ and the projective lightcone $\Sigma$.

We take the Cartesian coordinates of the lightcone to be $Y^A = (Y^\mu, Y^{d+1}, Y^{d+2})$, where $Y^{d+2}$ is the ``extra'' time direction. As usual, we can define advanced and retarded ``times'' as $Y^\pm = Y^{d+1}\pm Y^{d+2}$, which are null coordinates. In these coordinates we have
\begin{align}\label{eq:null_rays}
0 = \eta_{AB} Y^A Y^B
=  \eta_{\mu\nu} Y^\mu Y^\nu + (Y^{d+1})^2 - (Y^{d+2})^2
=  \eta_{\mu\nu} Y^\mu Y^\nu  + Y^+Y^-
\, .
\end{align}
A map $\Sigma \to M$, i.e., from the projective lightcone to $M$, is defined as
\begin{align}\label{eq:embedding_to_flat}
 Y^A \to x^\mu = \frac{Y^\mu}{Y^+} \, ,
\end{align}
where $x^\mu$ are the coordinates of a point $p \in M$. This mapping is consistent because elements of the same equivalence, $Y^A \sim \lambda Y^A$, map to the same $x^\mu$ in $M$. In other words, we can always choose a representative such that $Y^+=1$ by rescaling with $\lambda=Y^+$.

Lorentz transformations of the embedding space induce conformal transformations on the lightcone. To see this, start from differentiating Eq.~\eqref{eq:null_rays}, which gives
\begin{align}\label{eq:differential_lightcone}
2 \eta_{\mu\nu}  Y^\mu {\rm d} Y^\nu + {\rm d} Y^+Y^- +  Y^+{\rm d}Y^-
=
0
\, .
\end{align}
It can be combined with Eq.~\eqref{eq:embedding_to_flat} to show
\begin{align}
 \eta_{AB} {\rm d} Y^A {\rm d} Y^B = (Y^+)^2 \eta_{\mu\nu} {\rm d}x^\mu {\rm d}x^\nu \, ,
\end{align}
The action of any element of the Lorentz group $L^{A}{}_B$ in the embedding space, $Y^A\to Y^{\prime A}=L^{A}{}_B Y^B$, leaves $\eta_{AB} {\rm d}Y^A {\rm d}Y^B$ invariant, thus
\begin{align}
 (Y^{\prime +})^2 \eta_{\mu\nu} {\rm d}x^{\prime \mu} {\rm d}x^{\prime \nu} = (Y^+)^2 \eta_{\mu\nu} {\rm d}x^\mu {\rm d}x^\nu \, .
\end{align}
The above invariance can be interpreted as a conformal transformation
\begin{align}
 \eta_{\mu\nu} {\rm d}x^{\prime \mu} {\rm d}x^{\prime \nu} = \Omega(x)^2 \eta_{\mu\nu} {\rm d}x^\mu {\rm d}x^\nu \, , \qquad {\rm where}   \qquad    \Omega(x)= \frac{Y^+}{Y^{\prime +}} \, .
\end{align}
The precise relations among the generators can be found elsewhere, for example in Ref.~\cite{Rychkov:2016iqz}.

On the other hand, to each point on $M$ we can associated a unique element on the projective lightcone. Consider the map $M \to \Sigma$
defined by
\begin{align}\label{eq:flat_to_embedding}
x^\mu \to Y^A = (Y^\mu, Y^{+}, Y^{-}) = Y^+ \Bigl(x^\mu, 1,-x^2\Bigr)
\, ,
\end{align}
in which $Y^+$ is arbitrary, in agreement with the equivalence $Y^A \sim \lambda Y^A$. The case $Y^+=1$ is known as the \emph{isometric embedding} in $\tilde{M}$.
It is straightforward to see that $Y^A$ satisfy $Y^2=0$, but also $Y^+ x^2 = - Y^-$. Later, we will also need the same map in Cartesian coordinates, i.e., $x^\mu \to Y^A = (Y^\mu, Y^{d+1}, Y^{d+2}) = Y^+ \Bigl(x^\mu, \frac{1-x^2}{2},\frac{1+x^2}{2}\Bigr)$. 
Most importantly, notice that the two maps \eqref{eq:embedding_to_flat} and \eqref{eq:flat_to_embedding} are surjective and their combination gives the identity, so they are inverse to each other.

An important point is that these maps can be extended to fields \cite{Weinberg:2010fx,Costa:2011mg}, with appropriate conditions and constraints. Consider, for simplicity, the case of a scalar field $\varphi(x)$ on $M$, which is what we will need in the rest of the paper.  It is trivial to extend the field to the embedding space as $\varphi(x) \to \Phi(Y)=\varphi(x)$, where $Y^A$
are defined by \eqref{eq:flat_to_embedding} in the isometric embedding $Y^+=1$.
It is clear that, for consistency with the projective nature of the lightcone, we must require homogeneity of $\Phi$ on the lightcone
\begin{align}
\Phi(\lambda Y)= \lambda^{-\Delta}\Phi(Y) \, ,
\end{align}
where $\Delta$ can be identified with the scaling dimension of $\varphi$. Thus, the field $\Phi(Y)$ is completely determined from its value on the section $(x^\mu, 1,-x^2)$, which itself is determined by the point $x^\mu$ of $M$. In short, from any homogeneous scalar field $\Phi$ on $\Sigma$, we can define a conformally invariant scalar field $\varphi$ on $M$ as 
\begin{align}\label{eq:fields_embedding_to_flat}
 \varphi(x) = (Y^+)^\Delta \Phi(Y(x))
\, ,
\end{align}
as verified by choosing $\lambda = (Y^+)^{-1}$ and using $\Phi((Y^+)^{-1}Y^A)=\Phi((x^\mu, 1,-x^2)) \equiv \varphi(x)$.
Similar procedures apply to fields with nonzero spin, though additional requirements, such as orthogonality to the lightcone for vectors, are necessary to reduce the number of degrees of freedom going from $d+2$ to $d$ dimensions \cite{osborn-notes}.

Now we extend the embedding formalism to a neighborhood of the projective lightcone, which is crucial for generalizing the construction to an ambient geometry.
We identifying two special coordinates, $t=Y^+$, which is the advanced time on $\tilde{M}$ and a new coordinate $\rho$ that parametrizes deviations from the lightcone.\footnote{%
The index labels $t$ and $\rho$ are reserved in the entire paper to these two coordinates, so they will never be used as dummy indices.
In the math literature labels of $t$ and $\rho$ are often denoted with $0$ and $\infty$, respectively \cite{Fefferman:2007rka}. The coordinate basis is also generally presented in the order $(t,x,\rho)$, rather than $(t,\rho,x)$ as we do in the following.
}
We write Cartesian coordinates on $\tilde{M}$ close to $\Sigma$ as
\begin{align}
 X^A = (X^\mu, X^{d+1}, X^{d+2}) = t \left(x^\mu, \frac{1+2\rho-x^2}{2},\frac{1-2\rho+x^2}{2}\right) \,.
\end{align}
It should be clear that for $\rho=0$ we have that $X^A=Y^A$, so we approach the codimension-$2$ null hypersurface $\Sigma$, in fact $\rho$ determines
$X^2 = 2t^2 \rho$ and whether we are inside or outside the lightcone depends on it being positive or negative.
In the coordinates $(t,\rho,x^\mu)$ we have that the metric on $\tilde{M}$ becomes
\begin{align}\label{eq:embeddig_metric}
 \tilde{\eta}=\eta_{AB} \dd X^A \dd X^B
 =
 2\rho \dd t^2+2t \dd t\dd\rho+t^2 \eta_{\mu\nu}\dd x^\mu \dd x^\nu \, .
\end{align}
While the hypersurface $\rho=0$ selects the projective lightcone, we have that $t$ gives different transversal sections of it. The important property is that the vector $T=t\partial_t$ defines a \emph{homothety} of $\tilde{M}$, that is, the Lie-derivative of the metric is proportional to the metric itself ${\cal L}_T \tilde{\eta} = 2 \tilde{\eta}$, which relates to scale transformations on $M$. Furthermore, notice how the $d$-dimensional metric $\eta_{\mu\nu}$ appears
in the $(\mu\nu)$ components of $\tilde{\eta}$ up to a coefficient $t^2$ which is completely fixed by the requirement that $T=t\partial_t$ is a homothety.

We should now reconsider the problem of mapping fields from the embedding to the base flat space, given that we have extended our coordinates to the region of $\tilde{M}$ neighboring $\Sigma$. Consider now a general a scalar field $\Phi(t,\rho,x)$ in the embedding space $M$. Since we want to define $\Phi$ only on the lightcone, for which $\rho=0$, we can take it to be $\rho$-independent, which represents the simplest extension. Moreover, since $t$ represent a scale transformations on $M$, we require 
\begin{align}
\Phi(t,\rho,x)
=
t^{\Delta_\varphi} \varphi(x) \, ,
\end{align}
where we now refer to $\Delta_\varphi$ that will be interpreted as the Weyl weight, and here is just defined as the negative of the scaling dimension $\Delta$ in agreement with the expression given in Eq.~\eqref{eq:fields_embedding_to_flat} and the fact that we are working with scalar fields. The difference between $\Delta$ and $\Delta_\varphi$ changes by a constant value for operators with spin because, in that case, it is necessary to solder the Lorentz frame to the tangent bundle using vielbeins, which have weight $1$, but this distinction plays no role here because we always work with scalars.

We now have all the ingredients to construct a curved-space generalization of the embedding formalism, which becomes the ambient space. The construction comes with several caveats, that we explore in the remainder of this section.
First, generalize $M$ to be a $d$-dimensional \emph{curved} Riemannian manifold $M$, with coordinates $x^\mu$ and equipped with a metric $g_{\mu\nu}(x)$. We take the metric to be a representative of a class of conformally related metrics.
A simple heuristic or ``bottom-up'' way to arrive at the ambient space is to generalize \eqref{eq:embeddig_metric} with the replacement
\begin{align}
\eta_{\mu\nu}
\to
h_{\mu\nu}(x,\rho) \qquad {\rm such ~~ that} \qquad \left. h_{\mu\nu}(x,\rho) \right|_{\rho=0}=g_{\mu\nu}(x)
\, 
\end{align}
which implies an embedding of the curved $M$ into a $(d+2)$-dimensional curved space $\tilde{M}$, known as the ambient space.
The limit $\rho=0$ is oftentimes referred to as ``boundary'' in this context, borrowing the term from an equivalent limit in the AdS/CFT construction that we briefly touch later.
With the replacement, we arrive at the ambient metric
\begin{align}\label{eq:ambient_metric}
 \tilde{g} = \tilde{g}_{AB} \dd X^A \dd X^B
 =
 2\rho \dd t^2+ 2t \dd t \dd\rho + t^2 h_{\mu\nu}(x,\rho)\dd x^\mu \dd x^\nu
 \, ,
\end{align}
expressed in coordinates $X^A=(t,\rho,x^\mu)$.
By construction
the metric $\eqref{eq:ambient_metric}$ has the manifest homothety ${\cal L}_T \tilde{g}=2 \tilde{g}$ for $T=t\partial_t$, which can be proven easily because, in the coordinates $(t,\rho,x^\mu)$, we have that ${\cal L}_T \tilde{g}_{AB} = t \partial_t \tilde{g}_{AB} = 2 \tilde{g}_{AB}$. We also have that,
for each fixed pair $(t,x^\mu)$, the map $\rho \to (t,\rho,x^\mu)$ parametrizes a null geodesic, oftentimes referred to as ``the lightcone structure'' of $M$. These are two of the defining properties of an ambient space's metric.

For now, the tensor $h_{\mu\nu}(x,\rho)$, which could be interpreted as a $d$-dimensional $\rho$-dependent metric, is still undetermined.
The third and final requirement on the ambient space, which fixes the form of the components $h_{\mu\nu}(x,\rho)$, is that the ambient metric must be Ricci flat, i.e.,
\begin{align}\label{eq:ambient-ricci-flat}
 \tilde{R}_{AB}=0
 \qquad {\rm for ~~ small} ~~ \rho
 \,.
\end{align}
This last requirement is trivially satisfied everywhere for the flat embedding metric $\tilde{\eta}$, but holds with some caveats in the general case
that are important to keep in mind.
For odd $d$ it is always possible to solve for Ricci-flatness
condition in a recursive Taylor expansion in $\rho$ with the boundary condition $\left.h_{\mu\nu}(x,\rho) \right|_{\rho=0}=g_{\mu\nu}(x)$.
For even $d$, i.e., $d=2n$, the Taylor expansion fails and the coefficients are determined up to a certain order, meaning that the Taylor series should be repalced with a trans-series involving $\log(\rho)$ \cite{graham-hirachi}.
From the point of view of the Taylor expansion, this manifests through
\begin{align}
\tilde{R}_{AB}= {\cal O}(\rho^n)\,.
\end{align}
The failure is instigated by geometric ``obstructions'' to Weyl invariance and is intimately linked to the conformal anomaly, as we discuss later in this section.

%

\subsection{Ambient space, obstructions and holography}
\label{sect:ambient-holography}

From now on we refer to the manifold $\tilde{M}$ equipped with metric $\tilde{g}$ given in \eqref{eq:ambient_metric} as the ambient space of the manifold $M$ with metric $g$. 
Loosely speaking, diffeomorphisms and Weyl invariances are the curved space ``lifts'' of Lorentz and conformal invariance, respectively.
Thus, it might be expected that diffeomorphisms of the ambient $\tilde{M}$ induce Weyl and diffeomorphisms transformations upon projection to $M$ in the limit $\rho=0$, which corresponds to the projective lightcone $\Sigma$. This is clarified later in this section, while now we discuss the intrinsic geometry of $\tilde{M}$ in the neighbor of the lightcone.

Consider the ambient's Ricci-flatness condition \eqref{eq:ambient-ricci-flat} and general $d$.
Qualitatively and very informally, we now regard $d$ as an analitically continued parameter.
The flatness condition can be used to fix $h_{\mu\nu}(x,\rho)$ in a neighbor of the lightcone $\rho=0$ using the Taylor expansion
\begin{align}\label{eq:power_law_exp}
 h_{\mu\nu}(x,\rho) = g_{\mu\nu}(x) + \rho h^{(1)}{}_{\mu\nu}
 + \frac{1}{2} \rho^2 h^{(2)}{}_{\mu\nu} +\cdots
 = \sum_{p\geq 0} \frac{\rho^p}{p!} h^{(p)}{}_{\mu\nu}
 \,,
\end{align}
where $h^{(p)}{}_{\mu\nu} = (\partial_\rho)^p h_{\mu\nu}|_{\rho=0}$ and $h^{(0)}{}_{\mu\nu}=g_{\mu\nu}$. Ideally, the Ricci-flatness condition should allow us to determine each expansion tensor 
$h^{(p)}{}_{\mu\nu}$ in terms of lower-order ones recursively. Then each tensor would be expressed in terms of the curvature of the $d$-dimensional metric $g_{\mu\nu}$ being the boundary condition.

The important subtlety is that the expansion \eqref{eq:power_law_exp} is valid to all orders only for \emph{odd} dimensions $d$. In \emph{even} dimensions $d=2n$, 
the expansion \emph{fails} at order $n$. This failure is best seen by continuing $d$ and treating it as a variable, which reveals the nonanalytic behavior as poles in the dimensionality
\begin{align}
 h^{(n)}{}_{\mu\nu} \overset{d\to 2n}{\sim}  \frac{1}{d-2n} {\cal O}^{(n)}{}_{\mu\nu}+\cdots \, ,
\end{align}
where the dots hide terms that are finite in the limit $d\to 2n$.
The poles multiply an obstruction tensors ${\cal O}^{(n)}{}_{\mu\nu}={\cal O}[g]^{(n)}{}_{\mu\nu}$
that depend on the curvatures of the metric $g_{\mu\nu}$,
which we give in a moment.
In even dimension, $d=2n$,
the expansion \eqref{eq:power_law_exp} should be replaced by
the transseries
\begin{align}\label{eq:log_exp}
 h_{\mu\nu}(x,\rho) = g_{\mu\nu}(x) +\cdots
 +\rho^n \Bigl(
 h^{(n)}{}_{\mu\nu}
 + t^{(n)}{}_{\mu\nu} \log \rho
 \Bigr)
 +\cdots
\end{align}
where the logarithmic term is not determined by Ricci-flatness \cite{graham-hirachi}, but should be given as an additional boundary condition.\footnote{In holography, the unknown coefficient is associated to the holographic energy-momentum tensor, $t^{(n)}{}_{\mu\nu} \sim \langle T_{\mu\nu}\rangle $ \cite{Parisini:2022wkb,Parisini:2023nbd}.
}

In order to see the connection with holography, it is useful to introduce the coordinates $(s,r)$ defined by $\rho = -\frac{r^2}{2}$ and $t=\frac{s}{r}$. The the metric \eqref{eq:ambient_metric} in the new coordinates $X^A=(s,r,x^\mu)$ becomes
\begin{align}
 \tilde{g}
 =
 -\dd s^2+ \frac{s^2}{r^2} \Bigl( \dd r^2+ h_{\mu\nu}(x,\rho(r))\dd x^\mu \dd x^\nu \Bigr) \, .
\end{align}
It is easy to see that, locally in the coordinate $s$, we have that the $(d+1)$-dimensional hypersurfaces for constant-$s$
have induced metric proportional to
\begin{align}
 g_{d+1} \sim \dd r^2+ h_{\mu\nu}(x,\rho(r))\dd x^\mu \dd x^\nu\,.
\end{align}
A consequence of the Ricci-flatness condition is that
${\rm Ric}_{d+1} \sim g_{d+1}$ on the boundary, so, asymptotically, they approach AdS spaces.
The ambient space is thus referred to as an asymptotically local Anti-de Sitter space (ALAdS) \cite{Parisini:2022wkb,Parisini:2023nbd}.
Given the intimate connection with holography, the formalism that we are about to flesh out can then be used to compute the holographic anomaly (see Refs.~\cite{Bugini:2018def,Jia:2021hgy} and references within).

\subsection{Geometry of the ambient space}
\label{sect:ambient}

Now we explore the intrinsic geometry of $\tilde{M}$ in proximity of the projective lightcone.
In practical calculations, solving $\tilde{R}_{\mu\nu}=0$ order by order is sufficient to determine all the expansion tensors $h^{(p)}{}_{\mu\nu}$ while keeping $d$ arbitrary to check for possible poles leading to obstructions \cite{Manvelyan:2007tk,Jia:2023gmk}. This is the strategy adopted in the following computations.
From now on, for all ambient tensors components, we reserve the labels $t$ and $\rho$ to denote the corresponding coordinate components.

In coordinates $(t,\rho,x^\mu)$, the inverse ambient metric has components
\begin{align}
 \tilde{g}^{\rho\rho} = -2 \rho t^{-2} \qquad
 \tilde{g}^{\rho t} = t^{-1} \qquad
 \tilde{g}^{\mu\nu} = t^{-2} h^{\mu\nu}
\end{align}
where $h^{\mu\nu}=h^{\mu\nu}(x,\rho)$ are the inverse components of the $h_{\mu\nu}(x,\rho)$, which we regard as a ``metric'' at fixed $\rho$.
In fact, we use $h_{\mu\nu}$ and $h^{\mu\nu}$ to raise/lower Greek indices.
For convenience, we denote with $D_\mu$ the unique torsionless connection compatible with $h_{\mu\nu}$, which, by construction, interpolates with the Levi-Civita connection $\nabla_\mu$ of $g_{\mu\nu}$ in the limit $\rho=0$, i.e., $D_\mu|_{\rho=0} = \nabla_\mu$.
Up to symmetries of the indices, the nonzero components of the ambient Riemann tensor are
\begin{align}\label{eq:ambient_riemann}
 \tilde{R}_{\mu\nu}{}^\sigma{}_\xi
 & =
 R[h]_{\mu\nu}{}^\sigma{}_\xi
 -\delta_{[\mu}{}^\sigma h'{}_{\nu]\xi}
 + h_{\xi[\mu}h'{}_{\nu]}{}^\sigma 
 + \rho h'{}_{[\mu}{}^\sigma h'{}_{\nu]\xi}   \, ,
 \\
 \tilde{R}_{\mu\nu}{}^t{}_\sigma
 & =
 -t D_{[\mu} h'{}_{\nu]\sigma}  \, ,
 \\
 \tilde{R}_{\rho \mu}{}^t{}_\nu
 & =
 \frac{t}{4} h'{}_{\mu\xi} h'{}^\xi{}_\nu - \frac{t}{2} h''{}_{\mu\nu}  \, ,
\end{align}
where the primes indicate $\rho$ derivatives (e.g., $h'{}_{\mu\nu}=\partial_\rho h_{\mu\nu}$),
and $R[h]_{\mu\nu}{}^\sigma{}_\xi$ is the Riemann curvature of $D_\mu$.
All other components are either zero or can be derived from the above using the metric and the symmetries of the curvature tensor.
The nonzero components of the ambient Ricci tensor $\tilde{R}_{AB}\equiv \tilde{R}_{CA}{}^C{}_B$ are
\begin{align}
 \tilde{R}_{\mu\nu}
 & =
 R[h]_{\mu\nu}-\frac{d-2}{2} h'{}_{\mu\nu}-\frac{1}{2}h_{\mu\nu} h'{}_\sigma{}^\sigma
 -\rho h'{}_\mu{}^\sigma h'{}_{\sigma\nu}
 +\frac{1}{2}\rho h'{}_{\mu\nu} h'{}_\sigma{}^\sigma
 +\rho h''{}_{\mu\nu} \, , \label{eq:ricci_flatness_munu}
 \\
 \tilde{R}_{\rho\rho}
 & =
 \frac{1}{4}h'{}_{\mu\nu} h'{}^{\mu\nu}-\frac{1}{2}h''{}_\mu{}^\mu \, , \label{eq:ricci_flatness_rhorho}
 \\
 \tilde{R}_{\rho\mu}
 & =
 \frac{1}{2} D_\nu h'{}_{\mu}{}^\nu
 -\frac{1}{2} D_\mu h'{}_{\nu}{}^\nu \, , \label{eq:ricci_flatness_rhomu}
\end{align}
and the ambient curvature scalar $\tilde{R}\equiv \tilde{g}^{AB}\tilde{R}_{AB}$ is
\begin{align}
 t^2 \tilde{R}
 & =
 R[h]
 -(d-2) h'{}_{\mu}{}^\mu
 -\frac{3}{2} \rho h'{}_{\mu\nu} h'{}^{\mu\nu}
 +\frac{1}{2}\rho h'{}_{\mu}{}^\mu h'{}_{\nu}{}^\nu
 +2 \rho h''{}_{\mu}{}^\mu \, .
\end{align}
The curvatures of $D_{\mu}$ have been constructed using the metric $h_{\mu\nu}$, thus $R[h]_{\mu\nu} = R[h]_{\sigma\mu}{}^\sigma{}_\nu$ and $R[h]=h^{\mu\nu} R[h]_{\mu\nu}$, but they reduce to the curvatures of $\nabla_\mu$ in the limit $\rho=0$, e.g., $R[h]|_{\rho=0} = R$.

The tensors $h^{(p)}{}_{\mu\nu}$ appearing in Eq.~\eqref{eq:power_law_exp} can be obtained by solving Eqs.~\eqref{eq:ricci_flatness_munu}, \eqref{eq:ricci_flatness_rhorho}, and \eqref{eq:ricci_flatness_rhomu} to zero iteratively, and we express them in terms of the conformally covariant tensors given in Appendix~\ref{sect:conformal-tensors}.\footnote{%
A general strategy involves solving for the traces using \eqref{eq:ricci_flatness_rhorho}, and simplifying gradients using 
\eqref{eq:ricci_flatness_rhomu}, then \eqref{eq:ricci_flatness_munu}
can be used to deduce the coefficients $h^{(p)}{}_{\mu\nu}$ order-by-order in the most efficient way.
}
For example, at zeroth order in $\rho$, the equation $ \tilde{R}_{\mu\nu}=0$ becomes
\begin{align}
R_{\mu\nu}
-
\frac{d-2}{2}h^{(1)}{}_{\mu\nu}
-
\frac{1}{2}g_{\mu\nu}h^{(1)}{}_\mu{}^\mu
=
0 \, .
\end{align}
Taking the trace and substituting back the result, we determine
\begin{align}
 h^{(1)}{}_\mu{}^\mu = 2 \mathcal{J} \, ,
 \qquad
 h^{(1)}{}_{\mu\nu} = 2 K_{\mu\nu} \, ,
\end{align}
where $\mathcal{J}$ is the trace of the Schouten tensor $K_{\mu\nu}$ defined in Appendix~\ref{sect:conformal-tensors}. This solves completely the zeroth order in $\rho$ of $\tilde{R}_{AB}=0$, including $\tilde{R}_{\rho\rho}=0$.
The obstruction highlighted previously in Sect.~\ref{sect:ambient-holography} for $d=2$ ($n=1$) can be seen by recalling the definition of $K_{\mu\nu}=(R_{\mu\nu} - \mathcal{J} g_{\mu\nu})/(d-2)$, which displays the dimensional pole and the obstruction tensor ${\cal O}^{(1)}{}_{\mu\nu}\sim R_{\mu\nu}-R/(2(d-1))g_{\mu\nu}$.

Similarly, but with a bit more work, at the first order in $\rho$, we find the solution
\begin{align}
 h^{(2)}{}_\mu{}^\mu = 2 K_{\mu\nu} K^{\mu\nu} \, ,
 \qquad
 h^{(2)}{}_{\mu\nu} = -\frac{2}{(d-4)} B_{\mu\nu} +  2 K_{\mu\sigma} K^\sigma{}_{\nu} \, ,
\end{align}
displaying the obstruction $1/(d-4)$ in the form of the Bach tensor $B_{\mu\nu}$.\footnote{%
This obstruction is visible in the flat-space limit in computations such as those of Refs.~\cite{Osborn:2016bev,Stergiou:2022qqj}.
}
At the second order in $\rho$ we find
\begin{align}
 h^{(3)}{}_\mu{}^\mu = -\frac{8}{d-4} B_{\mu\nu} K^{\mu\nu}\,,
 \qquad
 h^{(3)}{}_{\mu\nu} = \frac{2}{(d-6)(d-4)} \Box B_{\mu\nu} + \cdots \, ,
\end{align}
where the dots hide several curvatures that are at least quadratic in the curvature tensors introduced above. We see that the obstruction in $d=6$ has the form $\Box B_{\mu\nu} +\cdots$.
Higher orders generalize this pattern, with the tensors $h^{(n)}{}_{\mu\nu}$
having poles in $1/(d-2n)$, multiplying obstructions of the form
$(\nabla^2)^{n-1}B_{\mu\nu}+\cdots$.
Again the dots hide terms that are at least quadratic in the lower order curvatures.

The obstruction in $d=2$ is somewhat special in that 
${\cal O}^{(1)}{}_{\mu\nu}|_{d=2}\sim R_{\mu\nu}-R/2g_{\mu\nu}=0$,
because $R_{\mu\nu}=R/2 g_{\mu\nu}$ in two dimensions.
As usual, the case $d=2$ is special because we should distinguish between global conformal and Virasoro invariances \cite{Nakayama:2016dby,Nakayama:2019xzz},
but we know that the obstruction still stands, which can be checked explicitly \cite{Zanusso:2023vkn}.
The same is not true for the other obstruction tensors, for example the Bach tensor can be nonzero in $d=4$.

There are two ways to look at the obstruction tensors.
It is a simple exercise, that we perform in more detail later,
to check that the Weyl transformations of the obstruction tensors are homogeneous in general $d$. If we multiply both sides of the equations defining the obstructions by the pole, this is telling us that
in each even dimension $d$ we have a ``new'' tensor that is conformally covariant, i.e., close to $d=4$ we have $2\sigma B_{\mu\nu}+\delta_\sigma B_{\mu\nu} \propto (d-4)(\cdots)$, where $\delta_\sigma$ is a Weyl rescaling $\delta_\sigma g_{\mu\nu}=2\sigma g_{\mu\nu}$. So in $d=4$, and only in $d=4$, we have a new conformally covariant tensor: the Bach tensor. This structure generalizes to all subsequent even dimensionalities \cite{Fefferman:2007rka}.
Another important point is that it can be shown that ${\cal O}^{(n)}{}_{\mu\nu} \sim (\nabla^2)^{n-1}B_{\mu\nu}+\cdots$, which implies that, even though we have new obstructions in higher dimensionalities,
such obstructions can all be expressed in terms of the Bach tensor.
This is a general fact discussed in relation to the energy-momentum operator of a conformally coupled theory being a primary operator for a specific coupling with $B_{\mu\nu}$ in Refs.~\cite{Stergiou:2022qqj,Osborn:2015rna}.

\subsection{Extrinsic geometry of the null $\rho=0$ hypersurface}\label{sect:extrinsic}

Any vector proportional to $\partial_t$ is tangent to the projective lightcone $\Sigma$, but it is also ``orthogonal'' to it because $\partial_t$ is null. The extrinsic geometry of timelike and spacelike hypersurfaces is generally captured by the first and second fundamental forms, which are induced metric and extrinsic curvature, but for null hypersurfaces some standard definitions are more complicate, because the induced metric of rank $(d-2)$ is degenerate from the point of view of the ambient space.

We begin by considering the homothety vector $T=t\partial_t$, which, in coordinates $X^A=(t,\rho,x^\mu)$ becomes $T^A = (t,0,0)$.
The specific choice of $T$ is not crucial for this construction as long as the chosen vector is null and tangent to the hypersurface, but this choice of $T$ is convenient to have homogeneously scaling tensors. Additionally,
we have that $T$ satisfies the autoparallel equation $T^A \tilde{\nabla}_A T^B=0$, with $\tilde{\nabla}$ the symmetric and metric compatible connection of $\tilde{g}$ in the ambient space, which is a general outcome of the lightcone structure.
To construct an orthonormal basis, we first introduce an auxiliary null vector field, denoted $P$, such that $\tilde{g}(P,T)=-1$. 
The vector $P$ can actually be chosen in various ways corresponding to alternative ways to construct the null geometry, but, if we also require that $P^\mu = 0$, we have that
$P^A = (0,-t^{-2},0)$ is the unique choice.
In other words, $P= -t^{-2}\partial_\rho$.
It is easy to verify that $P$ is also a solution of
the autoparallel equation $P^A \tilde{\nabla}_A P^B=0$.

A local nonorthonormal basis for the tangent space of the ambient manifold is completed by including additional $d$ ``spacelike'' vectors $V^A_{\mu}$,
where $\mu$ is a label, not a tensor index.
We choose them to be orthogonal to both $T$ and $P$:
$\tilde{g}(T,V_{\mu})=\tilde{g}(P,V_{\mu})=0$.
The chosen conventions impose
$V^A_{\mu} \overset{*}{=} (0,0,\delta^{\nu}_\mu)$, so $V_\mu = \partial_\mu$, up to a local transformation of the basis.
A transverse metric $\tilde{h}$ for $\Sigma$ replaces the first fundamental form of the standard construction. It is defined as
\begin{align}
 \tilde{h}_{AB}
 =
 \left.T_A P_B+ P_A T_B + \tilde{g}_{AB} \right|_{\rho=0} \, ,
\end{align}
and it is orthogonal to any linear combination of $T$ and $P$
by virtue of $\tilde{g}(T,P)=-1$.
The transverse metric is effectively the metric of a $d$-dimensional manifold, that is, the lightcone $\Sigma$. Using the coordinates $(t,\rho,x^\mu)$, in components it is
\begin{align}
 \tilde{h}_{AB}
 =
 \left(\begin{array}{ccc}
  0 & 0 & 0 \\
  0 & 0 & 0 \\
  0 & 0 & t^2 h_{\mu\nu}|_{\rho=0}
 \end{array}\right) \, ,
\end{align}
which suggests once more that $g_{\mu\nu}(x) = h_{\mu\nu}(x,\rho=0)$ is the natural metric on $\Sigma$, up to rescalings of $t$ that correspond to conformal transformations in the flat-space limit, as discussed previously. In other words, the intrinsic geometry of $\Sigma$ is described by the metric $g_{\mu\nu}$ up to the factor $t^2$ that depends on the embedding.

As for the extrinsic properties, these are naturally associated to ``normal'' derivatives with respect to the vectors $T$ and $P$. The Lie-derivative with respect to the vector $T$ is not particularly interesting because ${\cal L}_T \tilde{h}= 2\tilde{h}$, which is caused by the fact that $T$ is a homothety.
A geometrically interesting extrinsic curvature can be defined using the auxiliary vector $P$ as $\tilde{K}_{AB} \equiv {\cal L}_P \tilde{h}_{AB}= \tilde{\nabla}_{(A} P_{B)}$. Using the coordinates $(t,\rho,x^\mu)$, in components it is
\begin{align}
 \tilde{K}_{AB}
 =
 -\frac{1}{2}\left(\begin{array}{ccc}
  2t^{-2} & 0 & 0 \\
  0 & 0 & 0 \\
  0 & 0 &  h'{}_{\mu\nu}|_{\rho=0}
 \end{array}\right) \, .
\end{align}
The complete extrinsic information of a null hypersurface is typically captured by three quantities on $\Sigma$: a scalar, a vector, and a symmetric tensor, where the word tensor is meant from the point of view of the $d$-dimensional manifold \cite{JKC}. The latter plays the role of the second fundamental form in the standard nondegenerate case. In the ambient space, the projective lightcone only has the scalar and the tensor
\begin{align}
m &= \tilde{h}^{AB}\tilde{K}_{AB} = -\frac{1}{2 t^2} h^{\mu\nu}  h'{}_{\mu\nu}|_{\rho=0}
\,, \\
K_{\mu\nu} &= \tilde{K}_{AB} V^A_\mu V^B_\mu = \frac{1}{2} h'{}_{\mu\nu}|_{\rho=0} 
\,.
\end{align}
Inserting the previously determined $h^{(1)}{}_{\mu\nu}$ from the Ricci-flatness condition, it turns out that $h'{}_{\mu\nu}|_{\rho=0}=2K_{\mu\nu}$, where $K_{\mu\nu}$ is the Schouten tensor of the spacetime with metric $g_{\mu\nu}=h_{\mu\nu}|_{\rho=0}$ defined in Appendix~\ref{sect:conformal-tensors}. Interestingly, the degeneracy in the standard notation of the Schouten and the extrinsic curvature tensors -- both often denoted with the letter $K$ -- is harmless here, as they are in fact the same object. It is also easy to see that the scalar $m$ is related to its trace ${\cal J}$ and brings no new information.

\subsection{Ambient preserving PBH diffeomorphisms}
\label{sect:pbh}

A general diffeomorphism generated by an arbitrary vector $\zeta^A=\zeta^A(t,\rho,x)$ in the ambient space transforms the metric $\tilde{g}$ as
\begin{align}
 \delta_\zeta \tilde{g}_{AB}=
 ({\cal L}_\zeta \tilde{g})_{AB}
 = \zeta^C \partial_C \tilde{g}_{AB} + \tilde{g}_{AC} \partial_B \zeta^C
 + \tilde{g}_{BC} \partial_A \zeta^C \, ,
\end{align}
where $\mathcal{L}_\zeta$ denotes the Lie derivative with respect to $\zeta$. The PBH (Penrose-Brown-Henneaux) diffeomorphisms \cite{Brown:1986nw,Schwimmer:2000cu} are the subgroup of diffeomorphisms such that the properties of the ambient space do not change. In practice, the PBH diffeomorphisms are those that preserve the general form \eqref{eq:ambient_metric} of the metric, meaning that we must constrain
\begin{align}
 \delta_\zeta \tilde{g}_{tt}=\delta_\zeta \tilde{g}_{t\rho}=\delta_\zeta \tilde{g}_{\rho\rho}
 =\delta_\zeta \tilde{g}_{t\mu}=\delta_\zeta \tilde{g}_{\rho\mu}=0 \, .
\end{align}
On the other hand, the $(\mu\nu)$ component of the metric is allowed to change as long as it remains in the form $\tilde{g}_{\mu\nu}= t^2h_{\mu\nu}$ for some $h_{\mu\nu}(x,\rho)$. Together they ensure that the ambient metric \eqref{eq:ambient_metric} remains the in form same for some transformed $h_{\mu\nu}$.

The first four conditions imply the following equations for the components of the generator of the ambient diffeomorphism
\begin{align}
 &\zeta^\rho+t\partial_t \zeta^\rho+2\rho \partial_t \zeta^t=0 \, , \label{eq:PHB_eq1}
 \\
 &\zeta^t+2\rho \partial_\rho \zeta^t +t \partial_\rho \zeta^\rho+t \partial_t \zeta^t=0 \, , \label{eq:PHB_eq2}
 \\
 & t\partial_\rho \zeta^t=0 \, , \label{eq:PHB_eq3}
 \\
 & 2\rho \partial_\mu \zeta^t + t \partial_\mu \zeta^\rho + t^2 h_{\mu\nu}\partial_t \zeta^\nu=0 \, , \label{eq:PHB_eq4}
\end{align}
which heavily constrain the generator $\zeta^A$.
A solution of the previous system is easy to get: from Eq.~\eqref{eq:PHB_eq3} it follows $\zeta^t=\sigma_1(t,x)$, which, once inserted in Eq.~\eqref{eq:PHB_eq2}, yields $ \zeta^\rho=- \sigma_1\frac{\rho}{t}-\rho\partial_t\sigma_1$. Plugging both in Eq.~\eqref{eq:PHB_eq1} one obtains that $\sigma_1$ can be at most linear in $t$, i.e., $\partial^2_t \sigma_1=0$. Therefore, we write $\sigma_1 = t \sigma(x)$ and $\zeta^\rho = - 2\rho \sigma(x)$. In this way, from Eq.~\eqref{eq:PHB_eq4} it is easy to obtain that $\zeta^\mu$ is independent from $t$.
In conclusion, the solution must depend only on two functions $\zeta^A \to (\sigma(x),\xi^i(x,\rho))$ and reads
\begin{align}
 \zeta^t= t \sigma(x) \, ,
 \qquad
 \zeta^\rho =-2\rho \sigma(x) \, ,
 \qquad
 \zeta^\mu = \zeta^\mu(x,\rho) \, .
\end{align}

The final condition $\delta_\zeta \tilde{g}_{\rho\mu}=0$ is crucial in that it relates the two functions further. In components it implies
\begin{align}
 t \partial_\mu \zeta^t + t^2 h_{\mu\nu} \partial_\rho \zeta^\nu=0 \, ,
\end{align}
which, using the explicit parametrization of $\zeta^A$, becomes
\begin{align}
 h_{\mu\nu} \partial_\rho \zeta^\nu= -\partial_\mu \sigma \, .
\end{align}
We can integrate the relation to obtain
\begin{align}
\zeta^\mu(x,\rho) = \xi^\mu(x)-\int_0^\rho {\rm d}\rho' h^{\mu\nu}(x,\rho') 
\partial_\nu \sigma(x) \, ,
\end{align}
where we have chosen the boundary condition $\zeta^\mu(x,0)=\xi^\mu(x)$. We can determine the expansion of $\zeta^\mu$ in powers of $\rho$ as a function of $h_{\mu\nu}$ and $\sigma$. 
Introducing the expansion of $\zeta^\mu(x,\rho)$, with the initial condition $\zeta^\mu(x,0)=\zeta^{(0)\mu}=\xi^\mu(x)$, we have
\begin{align}
\zeta^\mu(x,\rho) = \xi^\mu(x)+ \rho \zeta^{(1)\mu} + \frac{1}{2}\rho^2 \zeta^{(2)\mu} +\cdots \, ,
\end{align}
and using the $\rho$-expansion of the inverse metric
\begin{align}\label{eq:rho_exp_inverse_h}
 h^{\mu\nu}=g^{\mu\nu} - \rho  h^{(1)}{}^{\mu\nu}  + \frac{1}{2} \rho^2 \left(  2  h^{(1)}{}^{\mu\gamma} h^{(1)}{}_{\gamma}{}^{\nu}  -  h^{(2)}{}^{\mu\nu}       \right) + \cdots \, ,
\end{align}
we can determine the coefficients at any order, e.g.,
\begin{align}
 \zeta^{(1) \mu} &= -\nabla^\mu \sigma \\
 \zeta^{(2) \mu} &=
 h^{(1) \mu}{}_\nu \nabla^\nu \sigma 
 \\
 \zeta^{(3) \mu} &= -  \left(  2  h^{(1)}{}^{\mu\gamma} h^{(1)}{}_{\gamma}{}_{\nu}  -  h^{(2)}{}^{\mu}{}_{\nu}       \right) \nabla^\nu \sigma 
\end{align}
Notice that the integral raises the power of $\rho$ by one.
Inserting the expansion tensor of the metric, which are determined by Ricci-flatness of the ambien space, to second order in $\rho$ we obtain
\begin{align}
\zeta^\mu = \xi^\mu -\rho \nabla^\mu \sigma  + \rho^2 K^\mu{}_\nu \nabla^\nu \sigma +\cdots \,.
\end{align}
Altogether, given that the expansion of $h_{\mu\nu}$ comes as a solution of the Ricci-flatness condition, the PBH diffeomorphisms are thus entirely determined by the pair of generators $(\sigma(x),\xi^\mu(x))$,
which suggests that they might be generating the group of spacetime conformal transformations \cite{Manvelyan:2007tk}. Here, by ``conformal transformations'' we mean the infinite-dimensional semidirect product of Weyl and diffeomorphisms groups, which gives the actual -- finite-dimensional -- conformal group as the subgroup of conformal isometries of $g_{\mu\nu}$.

\subsection{PBH diffeomorphisms as generators of conformal transformations}
\label{sect:pbh-vs-weyl}

There are some subtleties that are worth exploring in order to prove that the PBH diffeomorphisms on $\tilde{M}$ actually generate Weyl and diffeomorphisms transformations on $M$ upon projection to $\rho=0$.
To explore the relation with the conformal group of spacetime, notice that
\begin{align}\label{eq:tildePHB_diff}
 \delta_\zeta \tilde{g}_{\mu\nu} = 2t^2\Bigl(
  \sigma h_{\mu\nu}
 - \sigma \rho \partial_\rho h_{\mu\nu}
 + D_{(\mu} \zeta_{\nu)} 
 \Bigr) \neq 0 \, ,
\end{align}
where $D_\mu$ is the connection symmetric and compatible with $h_{\mu\nu}$ as before, and the index of $\zeta_\mu$ is lowered using $h_{\mu\nu}$, i.e., $\zeta_\mu= h_{\mu\nu}\zeta^\nu$. Thus, it admits the expansion
\begin{align}
\zeta_\mu
&=
\xi_\mu
+
\rho \left( h^{(1)}{}_{\mu\gamma} \xi^\gamma +    \zeta^{(1)}_{\mu}   \right)
+
\frac{\rho^2}{2}\left(    h^{(2)}{}_{\mu\gamma} \xi^\gamma + 2 h^{(1)}{}_{\mu\gamma} \zeta^{(1)}_{\gamma}     + \zeta^{(2)}_{\mu}             \right) +\cdots  \\ \nonumber
& \equiv
\xi_\mu
+
\rho \bar{\zeta}^{(1)}_{\mu}
+
\frac{\rho^2}{2} \bar{\zeta}^{(2)}_{\mu} +\cdots \, ,
\end{align}
with the understanding that $\zeta^{(p)}_{\mu} = g_{\mu\nu} \zeta^{(p)\nu}$. For convenience, let us denote collectively the difference between the two metrics as $h_{\mu\nu}=g_{\mu\nu}(x) + \sum_{p\geq 1} \frac{\rho^p}{p!} h^{(p)}{}_{\mu\nu} \equiv g_{\mu\nu} + H_{\mu\nu}$. Accordingly, we find the difference of their Levi-Civita connections
\begin{align}
\delta \Gamma^{\alpha}{}_{\nu\mu} = \frac{1}{2} g^{\alpha\lambda} \left(   \nabla_\nu H_{\mu\lambda} +  \nabla_\mu H_{\lambda\nu}  -  \nabla_\lambda H_{\nu\mu}            \right) + O(H^n)
\equiv
\rho  \delta \Gamma^{(1)\alpha}{}_{\nu\mu} 
+
\frac{\rho^2}{2} \delta \Gamma^{(2)\alpha}{}_{\nu\mu} +\cdots \, ,
\end{align}
where the number in the parenthesis stand for the power of $\rho$ hidden inside $H_{\mu\nu}$.\footnote{%
The notation $O(H^n)$ refers to higher powers of $H_{\mu\nu}$, which lead to an expansion in $\rho$ analogous to that in Eq.~\eqref{eq:rho_exp_inverse_h}. For instance, the first correction is of order $O(H^2)$ and can be represented schematically as $H\partial H + H^2 \partial g$, contributing to $\delta \Gamma^{(2)\alpha}{}_{\nu\mu}$ and higher orders. However, in the following discussion, we will confine our analysis to first order in $\rho$ and will therefore omit the explicit forms of these higher-order contributions, as they do not affect our results at this order.
}
Then, we have
\begin{align}
D_\mu \zeta_\nu
&=
\nabla_\mu \zeta_\nu - \delta\Gamma^{\alpha}{}_{\nu\mu}\zeta_\alpha
\nonumber \\
&=
\nabla_\mu \xi_\nu
+
\rho \left( \nabla_\mu \bar{\zeta}^{(1)}{}_\nu - \delta \Gamma^{(1)\alpha}{}_{\nu\mu}\xi_\alpha \right)  
+
\frac{\rho^2}{2} \left(   \nabla_\mu \bar{\zeta}^{(2)}{}_\nu   - 2  \delta \Gamma^{(1)\alpha}{}_{\nu\mu}   \bar{\zeta}^{(1)}{}_\alpha      -      \delta \Gamma^{(2)\alpha}{}_{\nu\mu}\xi_\alpha     \right) +\cdots \, ,
\end{align}
Notice that, thanks also to the projective structure of the lightcone with which we can rescale $t^2$, Eq.~\eqref{eq:tildePHB_diff} can be written as a genuine transformation of $h_{\mu\nu}$
\begin{align}\label{eq:PHB_diff}
 \delta_\zeta h_{\mu\nu} =
  2 \sigma (1-\rho\partial_\rho) h_{\mu\nu}
 + D_{\mu} \zeta_{\nu} 
 + D_{\nu} \zeta_{\mu} \, .
\end{align}

We can now expand Eq.~\eqref{eq:PHB_diff} in a Taylor series around $\rho=0$. For the ``boundary'' metric at $\rho=0$, we find the expected result
\begin{align}
 \delta_\zeta g_{\mu\nu}=\delta_{\sigma,\xi} g_{\mu\nu} = 
  2\sigma g_{\mu\nu}
 + \nabla_{\mu} \xi_{\nu} + \nabla_{\mu} \xi_{\nu} \, .
\end{align}
This shows that the transformation is a combination of a Weyl transformation, $\delta_\sigma g_{\mu\nu} = 2\sigma g_{\mu\nu}$ and a $d$-dimensional diffeomorphism $\delta_\xi g_{\mu\nu} = ({\cal L}_\xi g)_{\mu\nu}$. Notice once more that $D_\mu \rightarrow \nabla_\mu$ for $\rho \rightarrow 0$ for obvious reasons.
As we have already discussed, under general assumptions, invariance under the combination of Weyl and diffeomorphisms
implies conformal invariance under the conformal isometries of $g_{\mu\nu}$, assuming that the group is nonempty.
This is in agreement with the fact that the additional requirement
$\delta_\zeta\tilde{g}_{\mu\nu}=0$ selects the subgroup of PBH diffeomorphisms that leaves the ambient metric invariant, which is the ``true'' conformal group, i.e., the conformal isometries of $M$.

However, there is a tricky point which we find worth discussing further. 
The expansion of $\delta_\zeta h_{\mu\nu}$ in powers of $\rho$, induced by that of $h_{\mu\nu}$, should be consistent with the Ricci-flatness condition of the ambient space, which determines the expansion of $h_{\mu\nu}$ in terms of curvature tensors depending on $g_{\mu\nu}$.
For example, Eq.~\eqref{eq:PHB_diff} yields the transformation for the first order in $\rho$
\begin{align}
 \frac{1}{\rho}\delta_\zeta (\rho h^{(1)}_{\mu\nu})=
 \xi^{\rho}\nabla_\rho h^{(1)}_{\mu\nu} + h^{(1)}_{\mu\rho}\nabla_\nu \xi^\rho
 + h^{(1)}_{\nu\rho}\nabla_\mu \xi^\rho
 +\nabla_{\mu} \zeta^{(1)}_{\nu} + \nabla_{\mu} \zeta^{(1)}_{\nu}
 = {\cal L}_{\xi} h^{(1)}_{\mu\nu} + {\cal L}_{\zeta^{(1)}} g_{\mu\nu} \, .
\end{align}
Notice that, formally, $h^{(1)}_{\mu\nu}$ is not a tensor from the ambient point of view, because it obtained from a partial derivative with respect to the coordinate $\rho$, therefore the transformation $\delta_\zeta  h^{(1)}_{\mu\nu}$ is not an ambient covariant formula, but it should be covariant from the point of view of the manifold $M$ in which we have determined that $h^{(1)}_{\mu\nu}=2 K_{\mu\nu}$.
Nonetheless, by using the explicit form of the coefficients $\zeta^{(1)}_\mu= g_{\mu\nu}\zeta^{(1)\nu}$ and $h^{(1)}_{\mu\nu}$ and interpreting $ \delta_\zeta \equiv \delta_{\sigma,\xi}$, assuming $d>2$ we can write
\begin{align}
\delta_{\sigma,\xi} h^{(1)}_{\mu\nu}
 &= -2\nabla_\mu \partial_\nu \sigma + {\cal L}_{\xi} h^{(1)}_{\mu\nu} \, .
\end{align}
The two terms in the transformations should be interpreted as the Weyl and diffeomorphisms transformations of $h^{(1)}_{\mu\nu}$ and should also be consistent with the determination of $h^{(1)}_{\mu\nu}$ using Ricci-flatness given before.\footnote{%
In the literature pertaining the AdS/CFT correspondence the
$h^{(n)}_{\mu\nu}$ tensors have been obtained integrating their Weyl transformations
defined in this way, so the end result may differ by conformally covariant tensors starting from $n\geq 2$ \cite{Broccoli:2021icm}.
}
The latter transformation is obvious since there is the Lie derivative. As for the former,
we have that Ricci-flatness implies
$h^{(1)}_{\mu\nu}=2 K_{\mu\nu}$ and, given that the Shouten tensor transforms as $\delta_\sigma K_{\mu\nu} = -\nabla_\mu \partial_\nu \sigma$ under Weyl, the transformation of $h^{(1)}_{\mu\nu}$ is manifestly consistent with that of $2 K_{\mu\nu}$.
Similar checks can be performed for the next coefficients, confirming that the PBH diffeomorphisms are respectful of the underlying Ricci-flatness of the ambient geometry. Obviously, everything holds up to the obstruction caused by $\tilde{R}_{AB} \sim \rho^{n}$ in even dimensions $d=2n$.

%

\section{The integrable terms for the trace anomaly}
\label{sect:integrable-terms}

In this section we work out, using the ambient space formalism, the integrable terms that can be used to parametrize the anomaly.
The computations also highlight the Weyl-covariant differential operators that are helpful in the process of integrating the anomaly itself.

Before diving into the computations, it is useful to quickly recapitulate through one comprehensive example what we are looking for. We choose the $d=4$ trace anomaly because it brings forth all the important ingredients that appear in the general case. Consider a classical Weyl-invariant action $S[\varphi]$ of a Weyl-covariant field $\varphi$ coupled to a nondynamical metric.
The Noether identities tell us that on-shell the trace of the variational energy-momentum tensor, $T^{\mu\nu} = -\frac{2}{\sqrt{g}}\frac{\delta S}{\delta g_{\mu\nu}}$, is zero, i.e., $T^\mu{}_\mu=0$. Putting the action in a path-integral and integrating in $\varphi$ defines the effective action $\Gamma$, from which one can compute the expectation value
$\langle T^{\mu\nu} \rangle = -\frac{2}{\sqrt{g}}\frac{\delta \Gamma}{\delta g_{\mu\nu}}$, being $g_{\mu\nu}$ the source of the operator $T^{\mu\nu}$ (e.g., see Ref.~\cite{Hathrell:1981zb} for a self-interacting scalar). The path-integral induces an anomaly of Weyl invariance, known as trace anomaly \cite{Duff:1977ay}, which, at fixed points of the renormalization group, can be computed using the method outlined in Appendix~\ref{sect:KH_anomaly} and is of the form
\begin{align}
\langle T \rangle \equiv g^{\mu\nu}\langle T_{\mu\nu} \rangle = a E_4 + b W^2 + a' \Box_g R\,,
\end{align}
where $a$, $b$ and $a'$ are referred to as ``charges'' \cite{Cardy:1988cwa,Myers:2010xs}. We have that $E_4$ is the four-dimensional Euler's density and $W^2$ the square of Weyl tensor, and they are referred to as the $a$-anomaly and $b$-anomaly, respectively \cite{Deser:1993yx}. The $a$-anomaly is also known as the topological anomaly, for the obvious reason that it is related to a topological term, but also because the Weyl transformation behaves like $\delta_\sigma E_4 \propto (\nabla^2)^2 \sigma +\cdots $, where the term linear in $\sigma$ is related to a certain differential operator which plays a role in its subsequent integration \cite{Riegert:1984kt,Fradkin:1982xc}.
The anomaly includes also a total derivative of a vector, $\nabla_\mu J^\mu= a'\Box_g R$ with $J^\mu = a' \nabla^\mu R$, which can rather generally be eliminated with a scheme redefinition in the renormalization of $\Gamma$, given that $\delta_\sigma R^2 \propto \Box_g R$. For this reason, it is called a ``trivial'' anomaly or, in this case, $a'$-anomaly, where the charge $a'$ is scheme dependent \cite{Asorey:2006rm}.

The anomaly is constrained by a Wess-Zumino consistency condition $[\delta_\sigma,\delta_{\sigma'}]\Gamma=0$ \cite{Osborn:1991gm},
which can be structured as a cohomological problem \cite{Bonora:1983ff,Bonora:1985cq} or even more \cite{Jia:2023tki}.
The consistency is reflected in integrability conditions for the anomaly's constituents, that is,
$[\delta_\sigma,\delta_{\sigma'}] E_4 =0$ and $[\delta_\sigma,\delta_{\sigma'}] W^2 =0$.
The structure of $\langle T \rangle$ is general in that we expect that in any even dimension $d=2n$ the anomaly is the sum of three contributions, a topological $a$-anomaly, a Weyl covariant $b$-anomaly \cite{Cardy:1988cwa,Myers:2010xs}, and a trivial $a'$-anomaly which is a total derivative constrained by the Wess-Zumino consistency condition.\footnote{%
In $d=4$ the only trivial anomaly is $\Box_g R$ and satisfies the integrability condition, but in $d=6$ and higher total derivatives are not guaranteed to always come from Weyl variations of local terms \cite{Bastianelli:2000rs}.
}
All these charges have and have been regarded as potential generalizations of the central charge $c$ of Zamolodchikov's theorem \cite{Zamolodchikov:1986gt}, for example see Ref.~\cite{Duff:1993wm} for context and Refs.~\cite{Deser:1993yx,Cappelli:1990yc,Shore:1990wq,Osborn:1991gm,Anselmi:1999xk,Anselmi:1999uk} for examples, though the topological anomaly is generally regarded as the ``best'' candidate for irreversibility.

In the rest of this section we give a natural derivation of the candidates for $a$- and $b$-anomalies based on the ambient space. These are presented in Subsects.~\ref{sect:ambient-laplacian} and \ref{sect:invariants}, respectively.
The purpose is to shed light on their geometrical origin based on conformal geometry.
We do not give an ambient geometrical interpretation of the total derivative $a'$-anomalies, though we do often use bases of total derivatives for manipulations throughout the section.

\subsubsection*{Cautioning the Reader on the $d\to 2n$ limit}

A word of caution is necessary before going further.
We must explain how the obstructions are circumvented pragmatically and the point that we are about to make is an important
consideration that we will readdress in the conclusions. Assuming that we work out the expansion away from $d=2n$ even dimensions, in which the obstructions arise, the Taylor expansion in $\rho$ of the ambient space is perfectly fine.
Working in odd dimensions $d$, for example, would solve the problem, but
for the anomaly we need to be in even dimensions. Our point of view is that it is actually possible to ``analytically'' continue in $d$, as the expansions of the previous section clearly show. We use the notion of analytic continuation in a rather informal sense, implying that all formulas can be computed as functions of $d$, though some of them fail to work in even $d$ because of the poles. The important point is that
the subset of geometrical entities that are well-defined in the limit $d\to 2n$ preserves their geometric properties in the limit.
Among these are all the necessary ingredients to construct the $a$- and $b$-terms of the anomaly. In this sense, the limit $d\to 2n$ can circumvent the obstructions that we have observed in the construction of an ambient space. This is the framework that we are going to implicitly apply in the practical computations.

\subsection{Powers of the ambient Laplacian: GJMS hierarchy and $Q$-curvatures}
\label{sect:ambient-laplacian}

The implications of ambient diffeomorphisms inducing Weyl transformations introduced in Sect.~\ref{sect:ambient-main} extend beyond simple scalars. In fact, they affect other structures, such as ambient differential operators.
The ambient Laplacian is defined as $-\Box_{\tilde{g}}=-\tilde{\nabla}^2 \equiv-\tilde{g}^{AB}\tilde{\nabla}_A \tilde{\nabla}_B$, where $\tilde{\nabla}_A$ is again the unique torsionless covariant derivative compatible with $\tilde{g}$. On a general scalar field $\Phi= \Phi(t,x,\rho)$ in the ambient space, the Laplacian acts as
\begin{align}
 -\Box_{\tilde{g}} \Phi 
 =
 -
 \frac{1}{t^2} \Box_{h} \Phi
 -
 \frac{2}{t} \partial_t \partial_\rho \Phi
 +
 \frac{2\rho}{t^2} \partial^2_\rho \Phi
 -
h'{}_{\mu}{}^\mu  \left( \frac{1}{2t} \partial_t 
-
\frac{\rho}{t^2} \partial_\rho \right)\Phi 
 -
 \frac{d-2}{t^2} \partial_\rho \Phi
 \, .
\end{align}
From the considerations on the embedding formalism, we choose the field $\Phi$ of the scaling form
\begin{align}
 \Phi = t^{\Delta_\varphi} \varphi(x) \, .
\end{align}
A simple direct computation shows that the projection of $-\Box_{\tilde{g}} \Phi$ on the projective lightcone $\Sigma$ becomes
\begin{align}
 -\Box_{\tilde{g}} (t^{\Delta_\varphi} \varphi(x))|_{\rho=0}
 = t^{\Delta_\varphi-2}(-\Box_g \varphi - \Delta_{\varphi} \mathcal{J} \varphi)\, ,
\end{align}
where ${\cal J} = \frac{1}{2(d-1)}R$ is the trace of the Schouten tensor as before, which is defined in Appendix~\ref{sect:conformal-tensors}.
Thus, adopting the canonical dimension of a simple two-derivatives scalar field, $\Delta_{\varphi} = \frac{2-d}{2}$, then $\varphi$ is a conformal scalar on $M$
and the operation $t^{2-\Delta_\varphi} (-\Box_{\tilde{g}}) t^{\Delta_\varphi}= t^{-\frac{2+d}{2}} (-\Box_{\tilde{g}}) t^{\frac{2-d}{2}}$
identifies a conformally covariant operator
\begin{align}\label{eq:yamabe_operator}
 D_{2,d} \, \varphi(x)
 =
 t^{-\frac{2+d}{2}} (-\Box_{\tilde{g}}) (t^{\frac{d-2}{2}}\varphi)
 |_{\rho=0}
 =
 \left(-\Box_{g}+\frac{d-2}{4(d-1)} R\right) \, \varphi
 \,,
\end{align}
which is the well-known Weyl-covariant Yamabe operator. By construction,
${\cal L} \propto \sqrt{-g}\varphi D_{2,d} \varphi$ is a conformally invariant Lagrangian in any $d$, because it is the $t$-independent projection of $\sqrt{\tilde{g}}\Phi (-\Box_{\tilde{g}})\Phi$, except for $d=1$ in which case both Riemannian and conformal structures trivialize.
Notice that the endomorphism part of $D_{2,d}$ is proportional to $(d-2)$, so in $d=2$ we have that $D_{2,2} \equiv -\Box_g $
is a purely derivative operator, e.g., $D_{2,2} \varphi=0$ for constant $\varphi$.

The construction of Eq.~\eqref{eq:yamabe_operator} can be generalized to an arbitrary power of the ambient Laplacian \cite{GJMS,gover-hirachi}. We define the hierarchy of Graham-Jenne-Mason-Sparling (GJMS) operators
\begin{align}
 D_{2n,d} \varphi(x)
 \equiv
 t^{-\frac{2n+d}{2}} (-\Box_{\tilde{g}})^n (t^{\frac{2n-d}{2}}\varphi)
 |_{\rho=0}\, , \qquad \forall  n\in\mathbb{Z}^+
 \,.
\end{align}
By construction, they are conformally covariant operators acting on conformal ``higher-derivative'' fields with dimension $\Delta_\varphi=\frac{2n-d}{2}$ in any $d$, modulo potential obstructions
that we see in a moment \cite{gover-hirachi}. The ambient definition of the GJMS operators is quite compact, but their size increases rapidly
when expressed in terms of spacetime curvatures \cite{branson,branson-gover}.

A natural and useful way to deconstruct the $D_{2n,d}$ operators is
hinted from the $n=1$ case and consists in separating the derivative part from the constant one, which acts as an endomorphism. We write
\begin{align}\label{eq:GJMS_operators}
 D_{2n,d} 
 \equiv
 \Delta_{2n,d} + \frac{1}{2} (d-2n) Q_{2n,d}
\end{align}
where $\Delta_{2n,d}$ contains the derivative part, e.g., we have $\Delta_{2n,d}\varphi=0$ for constant $\varphi$. The above split defines implicitly the $Q$-curvatures
which are linear combinations with $2n$ derivatives constructed from Riemannian curvatures on $M$,
$Q_{2n,d} \sim {\cal R}^n$. It is important to realize that all objects, and especially the curvatures $Q_{2n,d}$, depend separately on both $n$ and $d$. Furthermore, both $Q_{2n,d}$ and $\Delta_{2n,d}$ are regular and nonzero in the limit $d\to 2n$,
which incidentally implies $D_{2n,2n}=\Delta_{2n,2n}$.
The important objects to keep in mind for the anomaly are precisely the
$d=2n$ limits, for which we adopt the notation
\begin{align}\label{eq:GJMS_operators_d2n}
 \Delta_{2n} \equiv \left.\Delta_{2n,d}\right|_{d=2n} = \left.D_{2n,d}\right|_{d=2n}\,,
 \qquad
 Q_{2n} \equiv \left.Q_{2n,d}\right|_{d=2n}
 = \lim_{d\to 2n} \frac{2}{d-2n} D_{2n} (1)
 \,.
\end{align}

We can determine the conformal properties of $D_{2n,d}$ either directly or indirectly, and thus deduce those of $\Delta_{2n}$ and $Q_{2n}$ as a result.
For example, by using $\delta_\sigma \int \dd^d x \sqrt{g} \varphi(x) D_{2n,d} \varphi(x)=0$, it is simple to obtain
the Weyl transformation
\begin{align}\label{eq:variation_Delta2n}
 \delta_\sigma D_{2n,d} (\dots) = -\frac{1}{2}(d+2n) \sigma D_{2n,d} (\dots) + \frac{1}{2} (d-2n) D_{2n,d} (\sigma \dots)\, ,
\end{align}
which should be interpreted as an operator expression where $D_{2n,d}$ acts on everything on its right, as the dots remind us. Technically speaking, the operators $D_{2n,d}$ have a Weyl-biweight, i.e., left- and right-Weyl-weights that can be read from Eq.~\eqref{eq:variation_Delta2n}. Clearly, Eq.~\eqref{eq:variation_Delta2n} can be written using Eq.~\eqref{eq:GJMS_operators} as
\begin{align}
\delta_\sigma D_{2n,d} 
=
-\frac{d+2n}{2} \sigma \Delta_{2n,d} -\frac{d^2-4n^2}{4} \sigma Q_{2n,d}
+
 \frac{d-2n}{2} \Delta_{2n,d} \sigma + \left( \frac{d-2n}{2} \right)^2\sigma Q_{2n,d} \, ,
\end{align}
where we omitted the dots for compactness. Since from the definition of Eq.~\eqref{eq:GJMS_operators} we have $\delta_\sigma D_{2n,d} \varphi =  \frac{1}{2} (d-2n)\delta_\sigma Q_{2n,d} \varphi$ on constant $\varphi$, we get by setting $\varphi=1$ the following transformation  
\begin{align}\label{conf_transf_Q}
 \delta_\sigma Q_{2n,d} = -2n \sigma Q_{2n,d} +\Delta_{2n,d} \sigma \, .
\end{align}

Specializing to the even dimensional case $d=2n$, the above conformal properties of the GJMS operators and $Q$-curvatures are the crucial ingredients for the integration of the conformal anomaly: evidently, $Q_{d}=Q_{d,d}$ satisfies the integrability condition
\begin{align}
 [\delta_\sigma,\delta_{\sigma'}]Q_{d}=0\,, \qquad
 \delta_\sigma Q_{d} = -d \sigma Q_{d} +\Delta_{d} \sigma
\end{align}
which is required for the parametrization of the anomaly and works in any even dimension. Furthermore, the operators $\Delta_{d} \sim (-\partial^2)^{d/2} +\cdots$ arise naturally when integrating the transformation of the $Q_{d}$ themselves.

The natural question is now if the $Q$-curvature $Q_{d}$ has anything to do with the Euler characteristic $E_{d}$ in even $d=2n$ dimensions.
A general theorem states that the integrated $Q$-curvatures are topological
up to conformal equivalence classes of metrics, $g_{\mu\nu} \sim g'_{\mu\nu}={\rm e}^{2 \sigma} g_{\mu\nu}$ \cite{branson,branson-gover,gover-peterson}, so they are conformal invariants. We state their relation more precisely later in Sect.~\ref{sect:anomaly-integration}, but, qualitatively, it means that
\begin{align}
 E_d = d \, Q_d + {\rm Weyl ~ invariants} + {\rm total~ derivative}
 \,,
\end{align}
so $Q_d$ is almost as topological as $E_d$, while having a simpler
transformation rule under Weyl rescalings. The usefulness is also in the fact that
the difference between $E_d$ and $Q_d$, up to a small normalization due to our conventions, can be completely understood as an unharmful linear redefinition of the charges of the anomaly. A comprehensive review of the mathematics can be found in Ref.~\cite{gover-peterson}.

The only subtlety in the above logic involves the total derivative, which does not manifestly satisfies the Wess-Zumino integrability condition. This issue is examined in more detail in Sect.~\ref{sect:anomaly-integration}. Nevertheless, we anticipate that the Weyl covariance properties of all the terms involved in the parametrization of the anomaly ensures that the total derivative term also inherently satisfies the Wess-Zumino consistency conditions.

\subsubsection{Examples of $Q$-curvatures and GJMS operators in $d=2,4,6$}

Let us show some explicit examples that can be computed algorithmically starting from the results of Sect.~\ref{sect:ambient-main} and using the projective definition \eqref{eq:GJMS_operators} of the GJMS operators.
The formulas for the differential operators $\Delta_{2n,d}$ below must be understood as operator relations in which derivatives act on everything on their right.

From the Yamabe operator \eqref{eq:yamabe_operator}, we see that in the $n=1$ case
\begin{align}
\Delta_{2,d} = -\Box_g\,,
\qquad
Q_{2,d} 
=
\frac{1}{2(d-1)} R
 = \mathcal{J} \, .
\end{align}
Repeating the computation for $n=2$, we obtain
\begin{align}
\Delta_{4,d} &=  \Box^2 + \nabla^\mu \left( 4 K_{\mu\nu} - (d-2)g_{\mu\nu} \mathcal{J} \right) \nabla^\nu \, ,\label{eq:Paneitz}\\
 Q_{4,d} &= 
  \frac{d}{2} \mathcal{J}^2 -2 K_{\mu\nu} K^{\mu\nu}-\Box \mathcal{J} \, ,\label{eq:Q4}
\end{align}
where the operator $\Delta_{4,d}$ is the well-known Paneitz operator \cite{paneitz} (also known as Fradkin-Tseytlin-Paneitz-Riegert operator \cite{Riegert:1984kt,Fradkin:1982xc} in $d=4$), expressed for arbitrary $d$. 
Finally, for $n=3$ we obtain
\begin{align}
\Delta_{6,d} &= - \Box^3 - 8 \left( \Box K_{\mu\nu} \nabla^\mu \nabla^\nu +  \nabla^\mu \nabla^\nu  K_{\mu\nu} \Box \right)
+
\frac{3(d-2)}{2} \Box \mathcal{J} \Box 
-2 \nabla^\mu \left( (10-d)	  \nabla_\mu \nabla_\nu  \mathcal{J}  +  \frac{8}{d-4} B_{\mu\nu} \right. \\ \nonumber   
&\left.+   24  K_{\mu\lambda} K_{\mu}{}^{\lambda}      -    4 (d-2)   K_{\mu\nu}  \mathcal{J}    \right) \nabla^\nu
+
4 \nabla^\nu  \left(   \Box  \mathcal{J}   + (d-4)    K_{\gamma\lambda} K^{\gamma\lambda}      - (\frac{3}{16} (d-2)^2  -  1 )       \mathcal{J}^2      \right)  \nabla_\nu
  \, ,\\
 Q_{6,d} &= 
 \frac{8(d-2)}{d-4} B_{\mu\nu} K^{\mu\nu} + \frac{(d-2)(d+2)}{4} \mathcal{J}^3
 -4(d+2)\mathcal{J} K_{\mu\nu} K^{\mu\nu}
 +8(d+2) K^{\mu\nu}K_{\mu\sigma} K_{\nu}{}^\sigma
 \\& \nonumber
 -16 K^{\mu\nu} K^{\sigma\theta} W_{\mu\sigma\nu\theta}
 -\frac{3d-2}{2} \mathcal{J}\Box\mathcal{J}
 -(d-6)\nabla_\mu\mathcal{J} \nabla^\mu\mathcal{J}
 +16 K^{\mu\nu}\nabla_\mu\nabla_\nu \mathcal{J}
 +8 \nabla_\mu K_{\nu\sigma}\nabla^\mu K^{\nu\sigma}
 +\Box^2 \mathcal{J} \, ,
\end{align}
where derivatives of $\Delta_{6,d}$ are understood as acting on everything on their right. The case $n=3$ was first covered by Branson, so we refer to the operator as Branson's operator \cite{branson,branson-gover}.
In the following, we will extensively use these $Q$-curvatures.

\subsubsection{Formulas in terms of the expansion tensors}

In the ambient space formalism it is natural to present these expression directly in terms of the expansion tensors $h^{(i)}$.
In the relations below, for convenience, the $h^{(i)}$ are treated as matrices
so that their contractions can be represented by traces and products, for example we have
${\rm tr}[h^{(i)}]=h^{(i)}{}_\mu{}^\mu$ and
$(h^{(i)}h^{(j)}){}_{\mu\nu}=h^{(i)}{}_{\mu\sigma}h^{(j)}{}^{\sigma}{}_{\nu}$. Traces and contractions are performed with $g_{\mu\nu}$ and its inverse.

The explicit expressions for $Q_{2n,d}$ show that $Q_{2n,d} \sim {\rm tr}[h^{(n)}]$, so they depend on the leading trace of the expansion tensor. This is important because the trace is sensitive to the pole $1/(d-2n-2)$, rather than the pole $1/(d-2n)$ that appears in the leading expansion tensor itself $h^{(n)}_{\mu\nu} \sim 1/(d-2n)$. This feature makes the $Q$-curvatures well-defined objects in the limit to even dimension, $d\to 2n$.
The $Q_{2,d}$ is simply
$Q_{2,d}
=
{\rm tr}[h^{(1)}]
$,
while $Q_{4,d}$ and $Q_{6,d}$ are more complicated and. They read
\begin{align}
 Q_4 = 
 -{\rm tr}[(h^{(1)}{})^2]
 +\frac{d}{8} ({\rm tr}[h^{(1)}])^2
 +{\rm tr}[h^{(2)}]
 -\frac{1}{2} \nabla^2 {\rm tr}[h^{(1)}]
 \,,
\end{align}
and
\begin{align}
Q_6 =&\nonumber\,
 8 {\rm tr}[(h^{(1)}{})^3]
 -d {\rm tr}[h^{(1)}] {\rm tr}[(h^{(1)})^2]
 +\frac{(d-2)(d+2)}{32} ({\rm tr}[h^{(1)}])^3
 -12 {\rm tr}[h^{(1)} h^{(2)}]
 +d {\rm tr}[h^{(1)}] {\rm tr}[h^{(2)}]
 \\ \nonumber&
 +4 {\rm tr}[h^{(3)}]
 +2 h^{(1)}{}^{\mu\nu} \nabla_\mu\nabla_\nu {\rm tr}[h^{(1)}]
 -2\nabla^2 {\rm tr}[h^{(2)}]
 -\frac{d-6}{4} \nabla_\mu {\rm tr}[h^{(1)}] \nabla^\mu {\rm tr}[h^{(1)}]
 + 4 {\rm tr}[h^{(1)}\nabla^2 h^{(1)}]
 \\ &
 -\frac{3d-2}{8} {\rm tr}[h^{(1)}]\nabla^2 {\rm tr}[h^{(1)}]
 +\frac{1}{2} \nabla^2\nabla^2 {\rm tr}[h^{(1)}]
 +4 \nabla_\mu h^{(1)}{}_{\nu\sigma}\nabla^\mu h^{(1)}{}^{\nu\sigma}  \, ,
\end{align}
respectively. 

The important ``structural'' observation is that the curvatures $Q_{2n}$ are sensitive, in principle, to all the obstruction poles in dimensions strictly less than $d=2n$ (not only in $d=2n-2$). In fact, they do exhibit poles in lower even dimensions $d=2n-2,2n-4,\cdots,2$ (but also $d=1$), highlighting the important aspect that the obstruction tensors are also obstructions to the lift from conformal to Weyl invariance in certain higher-derivative theories. For instance, while the higher-derivative generalized-free theory governed by the equation of motion $\Box^2\phi=0$ is a CFT in any dimension $d$, the Schouten tensor serves as a geometric obstruction to promoting this CFT to a Weyl-invariant theory in $d=2$. This insight suggests deeper connections with the structure of the corresponding CFT's spectrum \cite{Zanusso:2023vkn,Brust:2016gjy,Stergiou:2022qqj}.

\subsection{Conformal invariants in general dimension}\label{sect:invariants}

Conformally invariant/covariant scalar combinations of the curvatures
trivially satisfy the integrability condition of the conformal anomaly.
Here we give a very simple prescription to find such scalars \cite{Fefferman:2007rka}.

The procedure is based on the fact that ambient diffeomorphisms invariants become Weyl invariants when projecting them to the null hypersurface, and hence to $d$-dimensional base manifold $M$ \cite{Manvelyan:2007tk,Schwimmer:2000cu}. While obstruction tensors do arise in the ambient metric formalism for even dimensional conformal manifolds, they do not pose a fundamental problem for the construction of $d$-dimensional conformal invariants. Quite similarly to the case of $Q$-curvatures developed in Sect.~\ref{sect:ambient-laplacian}, this is because the construction relies on lower-order terms in the expansion, which are sufficient for the construction of Weyl invariants up to a mass dimension equal to the dimension of $M$ \cite{Fefferman:2007rka}. In any case, we adopt a pragmatic point of view by keeping $d$ arbitrary and explicitly checking that no poles arise.

To begin with, we should enumerate all scalars built with curvatures and covariant derivatives, schematically written
\begin{align}
\tilde{W}_{d,I}
=
{\rm tr}\,(
\tilde{\nabla}^{p_1}  \tilde{Riem} \times \dots \times \tilde{\nabla}^{p_r}  \tilde{Riem} 
) \, ,
\end{align}
such that $p_1+\dots+p_r + 2 m = d$, where $m$ is the number of Riemann tensor and $d$ the dimension of the boundary manifold.
Notice that, because of the Ricci-flatness of the ambient space, only the Riemann tensor contributes, so it is sufficient to enumerate only the structures the involve only the ambient Riemann tensor and neglect those that involve the Ricci and the Ricci scalar. For the same reason, no covariant derivative can be contracted with an index of the Riemann tensor it is acting upon, because, by Bianchi identities and Ricci-flatness, $\tilde{\nabla}^A \tilde{R}_{ABCD} =0 $.

The ambient diffeomorphisms invariants are projected to the null hypersurface trivially as
\begin{align}
W_{d,I} \equiv \tilde{W}_{d,I} |_{\rho=0,t=1}
\, ,
\end{align}
and the resulting objects are Weyl covariant in even $d$ dimensions by construction (they transform homogeneously with $\sigma$).
For convenience the projection uses the embedding $t=1$, because, given that the vector $T=t\partial_t$ is a homothety, we have that in general $\tilde{W}_{d,I} |_{\rho=0} = t^{-d} W_{d,I} $. Strictly speaking, the tensors $W_{d,I}$ are Weyl-covariant with dimension $d$, while their densitized versions are Weyl-invariant, $\sqrt{g} W_{d,I} \equiv \sqrt{\tilde{g}}\tilde{W}_{d,I} |_{\rho=0}$, and give invariant results when integrated.

To our knowledge, the number of elements $W_{d,I}$ as a function of $d$
is unknown, because it is a computationally hard problem. Instead the enumeration of the $W_{d,I}$s for lower $d$ is possible and relatively easy using modern software such as Mathematica's packages {\tt xAct}, {\tt xPert}, {\tt xTras} and {\tt Invar} \cite{xact,Brizuela:2008ra,Nutma:2013zea,Martin-Garcia:2007bqa},
with which we have also performed several checks.
The enumeration is rather direct, if the steps discussed above are followed, giving one Weyl-invariant on $M$ for each diffeomorphisms invariant on $\tilde{M}$, as long as Ricci-flatness and Bianchi identities are used when enumerating $\tilde{W}_{I,d}$. As far as we have checked, this happens also because there are no ``dimensional dependent'' identities among the curvatures that can be used to simplify the bases further. In any case, the important point is that bases can be provided for relatively high values of $d$, which go well beyond our needs in $d=2,4,6$.

\subsubsection{Examples of Weyl-invariants in $d=2,4,6$ and comments on $d=8$}

For simplicity we refer to the homogeneously transforming combinations are invariants, though they truly are invariants only if multiplied with $\sqrt{g}$.
As usual, invariants are constructed in general $d$ and then prolonged to the even dimensional case, such that the Taylor expansion of the ambient space is available.

In dimension $d=2$, the set of tensors $\tilde{W}_{2,I}$ is empty, because the only available scalar would be $\tilde{R}$, which is zero on the ambient space.
In dimension $d=4$, there is only one element,
which is $\tilde{W}_{4} = \tilde{R}_{ABCD} \tilde{R}^{ABCD}$.
We can project the invariant to the lightcone as follows
\begin{align}
 \tilde{W}_{4} |_{\rho=0,t=1} \equiv W_{4}
 = 
 W_{\mu\nu\alpha\beta} W^{\mu\nu\alpha\beta}
 \, ,
\end{align}
and the projection gives the unique conformal invariant in four dimensions.

The case $d=6$ is more interesting and only slightly more complicate. We should consider three independent invariants
\begin{align}\label{eq:ambient_scalars_6d}
\tilde{W}_{6,1}=\tilde{R}^{ABGR}\tilde{R}_{AMGN}\tilde{R}^{M}{}_{B}{}^{N}{}_{R}  \, ,         \quad\quad              
 \tilde{W}_{6,2}=\tilde{R}^{ABGR}\tilde{R}_{GRMN}\tilde{R}^{MN}{}_{AB}  \, ,           \quad\quad          
\tilde{W}_{6,3}=\tilde{ \nabla}_D     \tilde{R}^{ABGR}    \tilde{ \nabla}^D     \tilde{R}_{ABGR}
 \, .
\end{align} 
Notice that we do not need to take into account the scalar $ \tilde{R}^{ABGR} \tilde{\Box} \tilde{R}_{ABGR}$ because of the identity 
\begin{align}\label{eq:RiemBoxRiem}
\tilde{R}^{ABGR} \tilde{\Box} \tilde{R}_{ABGR}
&= -4 \tilde{W}_{6,1} - \tilde{W}_{6,2} \, ,
\end{align}
which can be proven using the Bianchi identities and Ricci-flatness.
We can verify that the ambient space easily yields the three known six dimensional conformal invariants, e.g., Refs.~\cite{Osborn:2015rna,Parker_6dinv}.
Upon projection, the first two scalars in \eqref{eq:ambient_scalars_6d} give 
\begin{align}\label{eq:6dinvariants_1_2}
W_{6,1}
=
W^{\alpha\beta\gamma\lambda}W_{\alpha\mu\gamma\nu}W^{\mu}{}_{\beta}{}^{\nu}{}_{\lambda}   \, ,         \quad\quad         
W_{6,2}
=
W^{\alpha\beta\gamma\lambda}W_{\gamma\lambda\mu\nu}W^{\mu\nu}{}_{\alpha\beta} \, ,
\end{align}
and the third scalar gives
\begin{align}
W_{6,3}
=
\nabla_\xi & W_{\mu\nu\alpha\beta} \nabla^\xi W^{\mu\nu\alpha\beta}
+
32 W_{\mu\nu\alpha\beta} \nabla^ \beta \nabla^\nu K^{\mu\alpha}
-
4(10-d) C^{\mu\alpha\beta}C_{\mu\alpha\beta}
+
16 K^{\mu\nu} W_{\mu}{}^{\xi\alpha\beta}  W_{\nu\xi\alpha\beta} 
\, .
\end{align}
It is a simple exercise to show that this basis coincides with the one of Ref.~\cite{Osborn:2015rna}, up to the linear redefinition of the element $W_{6,3}$. We have written $W_{6,3}$ in terms of the Cotton tensor, defined in Appendix~\ref{sect:conformal-tensors}.

In higher dimension, the strategy generalizes straightforwardly, although it is computationally more demanding.
To state the obvious, the incredible advantage is that enumerating diffeomorphisms invariants is
much simpler than enumerating conformal invariants, because diffeomorphisms invariance is manifest when writing covariant combinations of curvatures and covariant derivatives \cite{Fefferman:2007rka}.
The bases of diffeomorphisms invariant scalars with a given number of derivatives are known to a rather high order \cite{xact,Martin-Garcia:2007bqa,Nutma:2013zea}, and the only additional steps required to obtain Weyl invariants are to simplify multiple derivatives on Riemann tensors through Bianchi identities and eliminate combinations involving the Ricci tensor and the curvature scalar. In the end, it is always necessary to project to $\rho=0$, which may be computationally demanding, but not conceptually difficult.
Using the {\tt xAct} Mathematica suit \cite{xact,Martin-Garcia:2007bqa,Nutma:2013zea}, we have checked that in dimension $d=8$ we obtain precisely $12$ conformal invariants in agreement with the literature, see for example Ref.~\cite{Boulanger:2004zf}.

\section{Integration of the anomaly in any $d$}
\label{sect:anomaly-integration}

In this section we choose $d$ even, i.e., $d=2n$. Consequently, all the tensors coming from the ambient space construction are well-defined up to a certain order in $p$ of the expansion \eqref{eq:power_law_exp}. For the purpose of this section, it is sufficient that the objects $Q_d$ and $\Delta_d$, introduced in Sect.~\ref{sect:ambient-laplacian} as parts of the differential operators $D_{2n,d}$, and $W_{d,I}$, introduced in Sect.~\ref{sect:invariants}, are well-defined, which is in fact true and has been checked by explicit computations in $d=2,4,6$.
As hinted at the beginning of Sect. \ref{sect:integrable-terms}, the general form of the trace anomaly in $d=2n$ is expected to be the sum of three terms
\begin{align}\label{eq:general-anomaly-with-Euler}
 \langle T  \rangle 
 =
 \sum_I b_I \, W_{d,I} 
 +
 a E_{d}
 +
 \nabla_\mu J_d^\mu \,,
\end{align}
which are referred to as $b$-anomalies, $a$-anomaly and trivial (or $a'$-) anomalies, respectively \cite{Deser:1993yx,Duff:1993wm}. The terms of the anomaly come from the construction of a path-integral and depend on the conformal fields that are integrated-out. A relatively general procedure for obtaining $\langle T \rangle$ is sketched in Appendix~\ref{sect:KH_anomaly}, in which we also show a basic set of assumptions that lead to the desired result.

The $b$-anomalies are constructed from conformally covariant combinations which satisfy the integrability condition trivially. The possible conformal invariant densities have been constructed with the ambient space using the method discussed in Sect.~\ref{sect:invariants}, so we have parametrized them using the bases introduced there, which comes from projecting onto the lightcone the ambient diffeomorphisms invariants constructed from the ambient Riemann tensor.

The $a$-anomaly, also known as the topological anomaly, multiplies the Euler density, which we anticipate can be related to the $Q$-curvatures. In $d=2n$ it is defined as
\begin{align}
  E_{d}(g) = \frac{1}{2^n} R_{\mu_1\mu_2\nu_1\nu_2} \cdots R_{\mu_{2n-1}\mu_{2n}\nu_{2n-1}\nu_{2n}} \epsilon^{\mu_1\cdots\mu_{2n}}\epsilon^{\nu_1\cdots\nu_{2n}} 
  \,.
\end{align}
The charge $a$ of the topological anomaly is generally regarded as the most likely candidate for a generalization of Zamolodchikov's irreversibility theorem, and, in fact, it has been shown to be monotonic along the RG flow, although with some limitations, e.g., \cite{Shore:1990wq,Osborn:1991gm,Jack:1990eb,Komargodski:2011vj}.

Finally, the trivial anomaly $\nabla_\mu J_d^\mu$ is the total derivatives of some ``virial'' current \cite{Nakayama:2016dby}. Total derivatives are regarded as trivial because we assume that they can be integrated trivially. In fact, under the assumption that there exist a diffeomorphisms scale-invariant functional $F[g]$ such that
\begin{align}
 2 g_{\mu\nu} \frac{\delta}{\delta g_{\mu\nu}} F[g] = \sqrt{g}\, \nabla_\mu J_d^\mu
 \,,
\end{align}
it is sufficient to change the renormalization scheme of the effective action subtracting $\Gamma \to \Gamma' = \Gamma -F$ to obtain a new energy-momentum tensor that does not include the last term. This is of course possible only in a specific scheme and because the metric is not dynamical. More precisely, we assume that the trivial anomaly $ \nabla_\mu J_d^\mu$ is constrained to satisfy the Wess-Zumino consistency conditions, ensuring its integrability.

Notice that it is straightforward to prove that the variation of a scale-but-not-conformally invariant functional $F$ must be the total derivative of a certain vector \cite{Nakayama:2016dby}, while the opposite may not be obvious. The first nontrivial counterexample of a total derivative that is not integrable appears in $d=6$, as discussed in detailed in Ref.~\cite{Bastianelli:2000rs}. The analysis shows that there are seven distinct six-dimensional boundary terms can be constructed, but the space of trivial anomalies satisfying the consistency conditions is spanned by linear combinations of only six combinations. The explicit forms of these total derivatives are presented in Appendix~\ref{sect:basis6d}.
That said, in general there are multiple options for the functional $F[g]$ that integrates the total derivative given any $J^\mu_d$, which is another manifestation of the scheme dependence.

It is less obvious how to integrate the anomaly given in Eq.~\eqref{eq:general-anomaly-with-Euler} to obtain part of the effective action $\Gamma$ given that $\langle T \rangle = \frac{2}{\sqrt{g}}g_{\mu\nu}\frac{\delta \Gamma}{\delta g_{\mu\nu}}$. By ``integration of the anomaly'' we mean a general procedure such that it is possible to find a local and/or nonlocal action $\Gamma_{\rm an}$ such that
\begin{align}
 2 g_{\mu\nu} \frac{\delta}{\delta g_{\mu\nu}} \Gamma_{\rm an} = \sqrt{g} \,\langle T \rangle
 \,,
\end{align}
which ensures that the total effective action must be
\begin{align}
 \Gamma = \Gamma_{\rm conf} + \Gamma_{\rm an}\,,
\end{align}
where $\Gamma_{\rm conf}$ is a conformally (Weyl) invariant action, which could be local or nonlocal, that is not determined by integrating the anomaly.
While the lower dimensional case have been discussed extensively in the literature, the general version of the procedure in any even $d$ can be sketched rather straightforwardly using the ingredients of Sects.~\ref{sect:ambient-laplacian} and \ref{sect:invariants}.

To begin with, we use a general theorem that states that integrals of the $Q$-curvatures are topological invariants up to conformal classes of metrics \cite{gover-hirachi}. In practice, this means that there is a general relation
between the Euler density $E_d$ and the $Q_d$ curvature of the form
\begin{align}
 E_d = d \, Q_d + \sum_I e_I W_{d,I} +\nabla_\mu I_d^\mu
 \,,
\end{align}
where the coefficient $d$ in front of $Q_d$ is due to our normalization of the $Q$-curvatures. Inserting this expression in Eq.~\eqref{eq:general-anomaly-with-Euler} we find
\begin{align}\label{eq:general-anomaly-with-Q}
 \langle T  \rangle 
 =
 \sum_I \tilde{b}_I \, W_{d,I} 
 +
 \tilde{a} Q_{d}
 +
 \nabla_\mu \tilde{J}_d^\mu \,,
\end{align}
with new coefficients defined by $\tilde{a} = d\, a$, $\tilde{b}_I =b_I +e_I$ and $\tilde{J}_d^\mu = J_d^\mu +I_d^\mu$.

Let us begin with the integration of the third, which is the simplest term. As stated before, the total derivative term can in general be integrated trivially by finding any local functional $\Gamma_{\rm an, loc}$ such that
\begin{align}
 2 g_{\mu\nu} \frac{\delta}{\delta g_{\mu\nu}} \Gamma_{\rm an,loc} = \sqrt{g} \nabla_\mu \tilde{J}_d^\mu
 \,.
\end{align}
In general $\Gamma_{\rm an, loc}$ depends on the chosen scheme, inherited by the form of $J^\mu_d$. It is important to stress that this procedure is always applicable, as $ \nabla_\mu \tilde{J}_d^\mu$ is constrained to satisfy the Wess-Zumino consistency conditions. As shown in Appendix~\ref{sect:CC_Ed}, the Euler density naturally satisfies these conditions in arbitrary even dimensions. Additionally, $Q_d$ and $W_{d,I}$ trivially fulfill the consistency requirements, and we have assumed from the outset that $\nabla_\mu{J}_d^\mu$ also complies with them (being part of the anomaly). As a result, it directly follows that $\nabla_\mu I_d^\mu$ and $ \nabla_\mu \tilde{J}_d^\mu$ also satisfy the Wess-Zumino consistency conditions. To further support this point in Appendix~\ref{sect:basis6d} we explicitly verify in the first nontrivial case, which is $d=6$, that the appropriate $\nabla_\mu I_d^\mu$ can be derived as the variation of a local action.

In order to integrate the first two terms, let us introduce a fiducial metric $\overline{g}_{\mu\nu}$ representing the conformal class of $g_{\mu\nu}$ and parametrize $g_{\mu\nu}= {\rm e}^{2\sigma} \overline{g}_{\mu\nu}$. We refer to the functional that comes from the integration of the first two terms as $\Gamma_{\rm an, nl}$, where ``nl'' is for nonlocal as it will turn out to be nonlocal when written as a functional of $g_{\mu\nu}$. We have the relation
\begin{align}
 \frac{\delta}{\delta \sigma } \Gamma_{\rm an, nl}
 =
 \left.2 g_{\mu\nu} \frac{\delta}{\delta g_{\mu\nu}}\right|_{\overline{g}} \Gamma_{\rm an, nl} = \sqrt{g} \left(
 \sum_I \tilde{b}_I \, W_{d,I} 
 +
 \tilde{a} Q_{d}
 \right)
 \,.
\end{align}
We can write an ansatz for the functional $\Gamma_{\rm an, nl}$ interpreted as a separate function of $\sigma$ and $\overline{g}_{\mu\nu}$. Requiring that it is at most quadratic in $\sigma$, we have
\begin{align}
 \Gamma_{\rm an, nl}[\sigma, \overline{g}_{\mu\nu}]
 =
 \int {\rm d}^d x \sqrt{\overline{g}}\left(
 \sigma \sum_I \tilde{b}_I \, W_{d,I}[\overline{g}] 
 +
  \tilde{a} \left(
  \sigma Q_{d}[\overline{g}]
 +\frac{1}{2} \sigma \overline{\Delta}_{d} \sigma
 \right)
 \right)
 \,,
\end{align}
where $\overline{\Delta}_{d}$ is the operator ${\Delta}_{d}$ constructed with the fiducial metric. The relative coefficients of the terms are fixed by the requirement that it reproduces the first two terms of the anomaly. We can verify this explicitly by taking the functional derivative
\begin{align}
 \frac{\delta}{\delta \sigma }\Gamma_{\rm an, nl}
 =
 \sqrt{\overline{g}}\left(
 \sum_I \tilde{b}_I \, W_{d,I}[\overline{g}]
 +\tilde{a} \left(Q_{d}[\overline{g}]
 + \overline{\Delta}_d \sigma \right)
 \right)
 =
 \sqrt{g} \left(
 \sum_I \tilde{b}_I \, W_{d,I} 
 +
 \tilde{a} Q_{d}
 \right)
 \,,
\end{align}
where in the last step we have used the conformal properties of the integrable terms $\sqrt{\overline{g}}W_{d,I}[\overline{g}] = \sqrt{g} W_{d,I}$
and $\sqrt{\overline{g}}(Q_d[\overline{g}]+\overline{\Delta}_{d} \sigma) = \sqrt{g} Q_d$.

As anticipated, the label ``nl'' of $\Gamma_{\rm an, nl}$ refers to the fact that this term is nonlocal, if presented as a functional of the metric $g_{\mu\nu}$. To present it in a local form, we could regard the representative $\overline{g}_{\mu\nu}$ as the one picked by some ``gauge'' condition
\begin{equation}
 \chi[\overline{g}]=0\,,
\end{equation}
so that $\overline{g}_{\mu\nu}[g]$ is the representative of a given conformal orbit coming from the intersection with $\chi=0$ \cite{Barvinsky:2021ijq}.
Following and generalizing the presentation of Ref.~\cite{Barvinsky:2021ijq}, we can formally invert the gauge condition as
\begin{equation}
 \sigma = \Sigma_\chi[\overline{g}]\,,
\end{equation}
so that $g_{\mu\nu} = {\rm e}^{2 \Sigma_\chi[\overline{g}]} \overline{g}_{\mu\nu}$, which also tells us that $\Sigma_\chi[g]$ transforms by a shift under Weyl rescalings, i.e., $\Sigma_\chi[g] \to \Sigma_\chi[g] +\sigma$ for $g_{\mu\nu}\to {\rm e}^{2 \sigma} g_{\mu\nu}$.
Using this notion, we can write $\Gamma_{\rm an, nl}$ as a functional of a single metric, although gauge-dependent through $\Sigma_\chi$,
\begin{align}
 \Gamma_{\rm an, nl}[\Sigma_\chi[\overline{g}], \overline{g}_{\mu\nu}]
 =
 \int {\rm d}^d x \sqrt{\overline{g}}\Sigma_\chi[\overline{g}] \left(
  \sum_I \tilde{b}_I \, W_{d,I}[\overline{g}] 
 +
  \tilde{a} Q_{d}[\overline{g}]
 +\frac{\tilde{a}}{2} \overline{\Delta}_{d} \Sigma_\chi[\overline{g}]
 \right)
 \,.
\end{align}
Using once more the conformal properties of the integrable curvatures given in Sects.~\ref{sect:ambient-laplacian} and \ref{sect:invariants}, we have the gauge-dependent functional
\begin{align}
 \Gamma_{\rm an, nl}[g]
 =
 \int {\rm d}^d x \sqrt{g}\Sigma_\chi[g] \left(
  \sum_I \tilde{b}_I \, W_{d,I}[g] 
 +
  \tilde{a} Q_{d}[g]
 -\frac{\tilde{a}}{2} \Delta_{d} \Sigma_\chi[g]
 \right)
 \,.
\end{align}
This is a Liouville-like action, similar to the ones discussed in Ref.~\cite{Levy:2018bdc}, so we may call it the ``Liouville representation'' of the anomaly induced action.

There are multiple possible choices for the gauge-fixing, as many as those discussed in the $d=4$ case in Ref.~\cite{Barvinsky:2021ijq} (that we reconsider in Sect.~\ref{sect:d4} using entirely tensors derived from the ambient space) and certainly much more. The simplest gauge choice is probably $\chi[\overline{g}]=Q_d[\overline{g}]=0$, also seen in Ref.~\cite{Levy:2018bdc}, 
which generalizes the ``Riegert's'' choice and can be achieved for topologically simple spacetime
manifolds with a vanishing bulk part of the Euler characteristic as discussed in \cite{Riegert:1984kt,Barvinsky:2021ijq}. Then, we have
\begin{align}
 \Sigma_\chi[g] = \frac{1}{\Delta_d} Q_d\,,
\end{align}
where $\frac{1}{\Delta_d}$ is the Green function of $\Delta_d$, which gives nonlocal actions similar to Riegert's original manipulation in four dimensions. This choice has the potential problem of introducing $n$-th order poles, because $\frac{1}{\Delta_d} \sim \frac{1}{\Box^{d/2}}+\cdots$,
which may cause problems down the line, as discussed in Ref.~\cite{Coriano:2022jkn}, but we ignore them for now.

Thus, ignoring potential problems and combining everything together with the gauge choice, we have that the form of the anomaly constrains the effective action to be
\begin{align}
 \Gamma
 = \Gamma_{\rm conf}[g] + \Gamma_{\rm an,loc}[g]
 - \int {\rm d}^d x \sqrt{g} \left( \frac{\tilde{a}}{2} Q_d + \sum_I \tilde{b}_I W_{d,I} \right) \frac{1}{\Delta_d}  Q_d
 \,,
\end{align}
where the first term contains any possible local/nonlocal conformal action (undetermined by the anomaly), the second term is a scheme-dependent contribution, and the third nonlocal term is entirely constrained by the form of the trace anomaly and generalizes the Polyakov's and Riegert's actions \cite{Polyakov:1981rd,Riegert:1984kt}. The coefficients $\tilde{a}$ and $\tilde{b}_I$ can be given in terms of the original $a$ and $b_I$ as outlined for Eq.~\eqref{eq:general-anomaly-with-Q}.
The nonlocal part of the above action can be ``localized'' introducing at least two auxiliary fields \cite{Balbinot:1999ri}, however, if we exploit the fact that the conformally invariant part is undetermined, we can actually ``complete'' the square and write it as
\begin{align}
 \Gamma
 = \Gamma_{\rm conf}[g] + \Gamma_{\rm an,loc}[g]
 - \frac{1}{\tilde{a}}\int {\rm d}^d x \sqrt{g} \left( \frac{\tilde{a}}{2} Q_d + \sum_I \tilde{b}_I W_{d,I} \right) \frac{1}{\Delta_d}  \left( \frac{\tilde{a}}{2} Q_d + \sum_I \tilde{b}_I W_{d,I} \right)
 \,,
\end{align}
for a different $\Gamma_{\rm conf}[g]$, which only requires one auxiliary field to be localized \cite{Mottola:2006ew}.

In Sects.~\ref{sect:d2} and \ref{sect:d4} we provide examples of nonlocal actions in $d=2$ and $d=4$, respectively. Doing so we discuss vaguely the concept of ``ambiguities'', which may arise in the integration of the anomaly if different parametrizations of $\langle T\rangle$ in terms of $a$- and $b$-anomalies are chosen. We defer the case $d=6$ for Sect.~\ref{sect:d6}, after more appropriately having defined the concept of ambiguity at the beginning of Sect.~\ref{sect:ambiguities}.
Before going through with the first few simple examples of nonlocal action, in Sect.~\ref{sect:pondered} we remark the important role that the relation $\sqrt{\overline{g}}(Q_d[\overline{g}]+\overline{\Delta}_{d} \sigma) = \sqrt{g} Q_d$ has in the integration of the anomaly. This detour gives us the chance to provide a small generalization of the notion of ``pondered'' Euler density, introduced in Refs.~\cite{Anselmi:1999xk,Anselmi:1999uk}, which generalizes Riegert's approach to the integration.

\subsection{On the existence of a ``pondered'' Euler density}\label{sect:pondered}

Restricting our attention to conformally flat metrics $g_{\mu\nu}= {\rm e}^{2 \sigma} \eta_{\mu\nu}$, the pondered Euler density $\tilde{E}_d$ in $d=2n$ is defined as a minimal modification by a total derivative of a vector of the standard Euler density $E_{2n}$
\begin{align}\label{eq:pondered-def}
 \tilde{E}_{2n} = E_{2n} + \nabla_\mu V_{2n}^\mu
 \,,
\end{align}
such that
\begin{align}
 \sqrt{g} \tilde{E}_{2n} = {2n} (-\partial^2)^n \sigma
 \,,
\end{align}
where $\partial^2 = \eta^{\mu\nu}\partial_\mu\partial_\nu$ is flat-space's Laplacian (here implicitly expressed in Cartesian coordinates).
Notice that we use a slightly different normalization of both densities
to be consistent with the rest of this paper.
We can use the notion of $Q$-curvatures to ensure that such total derivative exists, in fact, it is straightforward to prove that
$Q_d \sim d \tilde{E}_d$ up to conformal invariant curvature terms that are zero for the conformally flat metric, i.e., $W_{\mu\nu\alpha\beta}=0$.
The pondered density has been used to ``solder'' the charge $a$ of the conformal anomaly to the charge $a'$ which can be related to a two-point function in $d=4$ and higher, for the purpose of circumventing the limitations of working with charges, such as $a$, that require higher than two-point functions in the quest for generalizations of Zamolodchikov's irreversibility theorem \cite{Anselmi:1999xk,Anselmi:1999uk}.

However, restricting the definition to conformally flat metrics has two main disadvantages. On the one hand we are not sensitive to conformal invariants constructed with the Weyl tensor and on the other hand the definition of $\tilde{E}_d$ itself may be ambiguous because $V^\mu_d$ is determined only up to conformal transformations.

Having already constructed the tower of GJMS operator, we can propose a general definition of pondered Euler density that works independently of the conformal class. Maintaining \eqref{eq:pondered-def}, but generalizing $g_{\mu\nu} = {\rm e}^{2\sigma} \overline{g}_{\mu\nu}$, the pondered Euler density is defined as the tensor that transforms
\begin{align}
 \sqrt{g} \tilde{E}_{2n}[g] = \sqrt{\overline{g}} \left(
 \tilde{E}_{2n}[\overline{g}] + {2n} \overline{\Delta}_{2n} \sigma
 \right)
 \,,
\end{align}
where $\overline{\Delta}_{2n}$ are the GJMS conformally covariant differential operators for the metric $\overline{g}_{\mu\nu}$.
Likewise the conformally flat case, it is straightforward to prove in general that such modification exists using the $Q$-curvatures.
In essence, the modification by a total derivative is made such that $\tilde{E}_d$ transforms homogeneously with respect to $\sigma$, while not altering most of the ``topological'' properties of the density itself, in the sense that it remains a total derivative.
These are precisely the properties that are needed when integrating the anomaly as shown in the previous section, so it is trivial to
repeat the same steps replacing $Q_d$ with $\tilde{E}_d$ to arrive at the nonlocal action.

In fact, the pondered Euler density can be seen as a generalization to any $d$ of Riegert's procedure \cite{Riegert:1984kt}. The limitation is that, outside conformally flat metrics, we have to introduce some sequence of conformally covariant operators, for which $\Delta_{2n}$ is the natural choice.
The advantage of working with the ambient space and the GJSM hierarchy of operators is that from the same object, that is, $D_{2n,d}$ in general $d$, one arrives \emph{at the same time} to the pair $(Q_d,\Delta_d)$ in $d=2n$
that enjoys a combined transformation as shown in Sect.~\ref{sect:integrable-terms}.

\subsection{Nonlocal action in $d=2$}\label{sect:d2}

In $d=2$ ($n=1$), the problem of structuring and integrating the anomaly is straightforward even without considering the ambient space construction,
but we stick to the ambient strategy for the presentation. In this dimension, there is no $b$-anomaly because the only ambient scalar that can be constructed is zero, $\tilde{R}=0$. There are also no derivative terms, since it is not possible to write down a dimension one covariant vector using only the metric (it would be possible including other nondynamical degrees of freedom such as torsion or nonmetricity \cite{Paci:2024ohq,Zanusso:2023vkn}). The first GJMS ambient Laplacianis the Yamabe operator \eqref{eq:yamabe_operator}
$D_{2,d}
=
-\Box_g
+
\frac{d-2}{4(d-1)}R
$,
from which we read $\Delta_2=-\Box_g$ and $Q_2=\frac{R}{2}$ in $d=2$. The $a$-anomaly is related to $Q_2$ trivially as $E_2=\tilde{E}_2=R$, so the two densities coincide.
We express the anomaly as 
\begin{align}\label{anomaly-d=2}
\langle T  \rangle 
=
a R
=
a \tilde{E}_2
=
\tilde{a} Q_2 
\, ,
\end{align}
where the $a$ depends on the field content and $\tilde{a}=2a$. For a free conformal scalar field with two derivatives, we have $a=\frac{1}{24\pi}$, which relates to the central charge being one $c=1$. In general, if a $2d$ CFT has central charge $c$, we have the relation $a=\frac{c}{24\pi}$.
Following the steps discussed in Sect.~\ref{sect:anomaly-integration}
we have a Liouville-like action
\begin{align}\label{local_EA_2d}
\Gamma[g,\sigma]=
a\int \dd^2 x\sqrt{{g}} \left( 
2\sigma {Q}_2 - \sigma {\Delta}_2 \sigma
\right)\,,
\end{align}
and, using the gauge condition $\sigma_\chi = \frac{1}{\Delta_2} Q_2$,
we integrate to the Polyakov action
\begin{align}\label{nonlocal_EA_2d}
\Gamma_{\rm nl}[g]=
a\int \dd^2 x\sqrt{g} \left( 
Q_2 \frac{1}{\Delta_2} Q_2
\right)
=
\frac{a}{4}\int \dd^2 x\sqrt{g} \left( 
R \frac{1}{-\nabla^2} R
\right).
\end{align}
It is simple to verify that for $a=c/(24\pi)$ the coefficient becomes
$a/4=c/(96\pi)$ in agreement with the standard result \cite{Mukhanov:2007zz}.
In $d=2$, there are no ambiguities in the process of constructing and integrating the anomaly. 

%

\subsection{Nonlocal action in $d=4$}\label{sect:d4}

In $d=4$ ($n=2$), the analysis is a bit more intricate.
The nonzero ambient scalar $\tilde{R}_{ABCD}^2$ projects to $W^2=W_{\mu\nu\sigma\lambda}^2$ in $\rho=0$ and $t=1$.
We thus have a genuine $b$-anomaly, $\langle T  \rangle_b \equiv \tilde{b} W^2$.
The charge of the $b$-anomaly is shown to be constrained for unitary CFTs \cite{Deser:1974cz,Cappelli:1990yc}.
From the second operator of the GJMS construction $D_{4,d}$ we read in $d=4$
\begin{align}
 \Delta_4 = -\Box^2 + 2 \nabla^\mu \left( R_{\mu\nu} - \frac{R}{3}g_{\mu\nu}  \right) \nabla^\nu \,,
 \qquad
 Q_4= \frac{1}{4}\Bigl(
 E_4- W^2 -\frac{2}{3} \Box R
 \Bigr)\,,
\end{align}
so we have the $a$-anomaly, which we parametrize as $\langle T  \rangle_a \equiv \tilde{a} Q_4$. Finally, thanks to the Bianchi identities, the only trivial $a'$-anomaly can be given by the total derivative
$\langle T  \rangle_{a'} \equiv a'  \nabla_\mu \tilde{J}^\mu$ with $\tilde{J}^\mu = \tilde{a}'\nabla^\mu R$ \cite{Asorey:2003uf,Asorey:2006rm}.
Combining the three contributions
\begin{align}\label{Qanomaly4d}
\langle T  \rangle 
=
\tilde{b} W^2
+
\tilde{a} Q_4
+
\tilde{a}'  \Box R
\,.
\end{align}
We can compare this parametrization with the more standard form
\begin{align}\label{anomaly-d=4}
\langle T \rangle 
=
b W^2
+
a E_4
+
a'  \Box R
\,,
\end{align}
to find the linear relations among the charges $(a,b,a')$ and $(\tilde{a},\tilde{b},\tilde{a}')$ of the two parametrizations:
$\tilde{b}=b+a$, $\tilde{a}=4a$ and $\tilde{a}'=a' + 2/3a$.
It is also quite simple to determine the relation between pondered Euler density and the standard Euler density \cite{Riegert:1984kt}
\begin{align}\label{pond_E4}
\tilde{E}_{4} 
=
E_4 
+
\nabla_\mu V^\mu_4 \,,
\qquad 
V_4^\mu 
= 
-\frac{2}{3} \nabla^\mu R\, .
\end{align}
Given that the total derivative of $V^\mu_4$ does not depend on the Riemann tensor, we have that
$\tilde{E}_{4}$ transforms linearly in the conformal factor $\sqrt{g'} \tilde{E'}_4
=
\sqrt{g} \left(
 \tilde{{E}}_4 + 4 {\Delta}_4 \sigma
 \right)
 $,
where $\Delta_4$ is the Paneitz operator,
so the original definition of the pondered Euler density and our generalization coincide for $d=4$.

The integrability conditions imply that there is no $R^2$ that appears independently in either of the above parametrization. This can be checked through explicit computations that, for classically conformally covariant fields, give
\begin{align}
(4\pi)^2\,b&=\frac{1}{120} N_\phi + \frac{1}{10} N_A - \frac{1}{20} N_\psi     \, \\
(4\pi)^2\,a&=-\frac{1}{360}  N_\phi - \frac{31}{180} N_A - \frac{11}{360} N_\psi  \, \\
(4\pi)^2\,a'&=\frac{1}{180} N_\phi - \frac{1}{10} N_A + \frac{1}{30} N_\psi  \,  ,   
\end{align}
where $N_\phi$, $N_A$ and $N_\psi$ denote the number of scalars, vectors\footnote{%
Maxwell's Abelian gauge theory is conformal in $d=4$, so the contribution of $N_A$ includes the ghosts. In $d\neq4$ the gauge-field action is incompatible with Weyl invariance for standard two-derivative actions, in which case the number of degrees of freedom changes discontinuously and there is only one ``ghost contribution'' as in Proca theories.
}
and $4d$ Dirac spinors, respectively, while there is no charge associated to $R^2$ by itself.

As for the integration of the various terms, we proceed starting from the simplest.
The trivial anomaly can be obtained as the variation of any square of local curvature, so we must choose one: $\Gamma_{\rm loc}[g]=\frac{a'}{12}\int \dd^4x \sqrt{g} R^2$, which is such that $-\frac{2}{\sqrt{g} }     g_{\mu\nu}    \frac{\delta}{\delta g_{\mu\nu}}    \Gamma_{\rm loc}[g] = a' \Box_g R$. This task can be accomplished equally well with the other
curvature square monomials and even their linear combinations, as long as the combination is \emph{not} one of the other anomalies. This is best understood as a cohomological problem of the trivial cocycle \cite{Bonora:1983ff,Bonora:1985cq}.

There are two ``popular'' options for the integration of the nonlocal part of the action \cite{Barvinsky:2021ijq}. The choice $Q_4[\overline{g}]=0$ discussed at the beginning of Sect.~\ref{sect:anomaly-integration} would be
\begin{align}
\Sigma_\chi = \frac{1}{\Delta_4}Q_4
\,,
\end{align}
and it is essentially the Riegert-Fradkin-Tseytlin gauge of Ref.~\cite{Barvinsky:2021ijq} in our framework and notation. It gives Riegert's original result \cite{Riegert:1984kt},
\begin{align}\label{eq:Riegert_Q_curvature}
\Gamma[g]
=
\Gamma_c[g]
-
\frac{\tilde{a}'}{12} \int \dd^4x\, \sqrt{g} \, 
R^2
-  \int \dd^4x\, \sqrt{g} \left(
\tilde{b} W^2  + \frac{\tilde{a}}{2}  Q_4  
\right)
\frac{1}{\Delta_4} Q_4
\,,
\end{align}
although written in a slightly different notation.
An alternative option would be $Q_{2,4}[\overline{g}]=0$, that we call Fradkin-Vilkovisky gauge \cite{Barvinsky:2021ijq}, which, from the ambient space point of view, is essentially a lift of the choice made in $d=2$ using the $d$-dependence of ambient operators and tensors
\begin{align}\label{eq:FBV_gauge}
\Sigma_\chi = -\log\left(1-\frac{1}{D_{2,4}}Q_{2,4}\right) = -\log\left(1+\frac{1}{\Box_g - \frac{R}{6}} \, \frac{R}{6}\right)
\,,
\end{align}
that is obtained assuming ${\rm e}^{-\Sigma_\chi} \to 1$ in the asymptotic region $\left|x\right|\to +\infty$ \cite{Fradkin:1978yw}.
The second choice gives a nonlocal action that is compatible with the results by Barvinsky \& Vilkovisky based on covariant perturbation theory \cite{Barvinsky:2021ijq} and the nonlocal heat kernel expansion \cite{Barvinsky:1987uw,Barvinsky:1990up,Codello:2012kq}. We do not write down the full nonlocal action in the second gauge because it is quite more notationally complicate than the previous one, but notice the particular structure that involves a logarithm as a consequence of gauge-fixing the conformal group.\footnote{%
An essentially similar problem was dealt with in Ref.~\cite{Martini:2024tie} when constructing group-like substructures of the Weyl group
and, from those results, it is natural to conjecture that there exist generalizations of the Fradkin-Vilkovisky gauge in dimensions higher than four (e.g., explicit examples are discussed for $d=6,8$ using lower-dimensional $Q$-curvatures).
These group-like substructures are known to lead to general nonlocalities \cite{Glavan:2024fsu}, which is precisely what we observe when integrating the anomaly.
On this note, given that in Ref.~\cite{Martini:2024tie}
the gauge fixings are interepreted as ``partial'' gauge fixings of Weyl invariance, it is natural to foresee the existence of a Gribov ambiguities problem.
}

There are two potential sources of ambiguities in the integration of the anomaly, one ``true'' and one ``fake''. A true ambiguity is related to the fact that we have multiple options to integrate the term $\Box R$ in the anomaly, since almost all squares of the curvatures give a term proportional to $\Box R$ upon variation. This true ambiguity is generally bypassed because it can be interpreted as a change of scheme, in that it can be subtracted from the full effective action with a redefinition of the renormalized interaction. A fake ambiguity is present because any linear ``redefinition'' $Q_4 \to Q_4 +c W^2$ or $E_4 \to E_4 + c W^2$ for some constant $c$ leads to terms that can be integrated equally well (in fact this is precisely the difference between the use of $\tilde{E}_4$ and $Q_4$), but this corresponds only to a change of basis, and the form of the nonlocal action remains invariant
because the combination $\tilde{b} \, W^2  + \frac{1}{2} {\tilde{a}} Q_4  $ is invariant.

\section{Generalized GJMS and integration ambiguities}\label{sect:ambiguities}\label{sect:new_GJMS}

As explained in detail in Sect.~\ref{sect:ambient-main} and applied in Sect.~\ref{sect:integrable-terms}, the projection of the ambient space Laplacians $D_{2n,d}$ gives us a natural pair $(\Delta_d,Q_d)$ of conformally covariant operators and $Q$-curvatures in $d=2n$ that allows to easily integrate the anomaly in even dimension. However, the choice of the ambient Laplacians for the projection is somewhat arbitrary, although arguably natural in its simplicity. In some cases, the arbitrariness is reflected in harmless shifts of the definition of $Q$-curvatures, e.g., like in the use of the pondered Euler density, which ultimately only reparametrizes the anomaly and does not change the final result for the nonlocal action. However, it has been observed before that in $d = 6$ there can be parametric families
of pairs that transform precisely like $(\Delta_6, Q_6)$ \cite{Ferreira:2017wqz,Ferreira:2018utt,Hamada:2000me}, meaning that some parameters of the nonlocal action that come from the integration of the anomaly are undetermined, leading to potential ambiguities. 

While we will not consider this possibility, as it is not directly related to the trace anomaly, it is worth noting that the GJMS construction can be generalized to include fields with spin. For instance, Ref.~\cite{Aros:2022ecb} illustrates such a generalization by deriving higher-derivative conformal operators acting on transverse-traceless symmetric rank-2 tensors and vectors.

In this section we want to give an ambient geometrical interpretation of these and other potential ambiguities which come from the choice of the pair $(\Delta_d,Q_d)$. We show how to generalize the sequence of GJMS ambient operators $(-\Box_{\tilde{g}})^n $ in a way that leads to potential ambiguities. We then classify them as either ``fake'' or ``true'', depending on their effect on the nonlocal action. True ambiguities are going to be those that actually make the nonlocal action different by introducing some parameters in it. It will be easy to see that true ambiguities appear only for $d \geq 6$, leading to parameters-dependent actions, while fake ones are present for $d \geq 4$, and correspond to innocuous reparametrizations of the anomaly coefficients.

Let us start by considering a modification of the ambient Laplacians in the form of some endomorphism
\begin{align}
(-\Box_{\tilde{g}})^n + \tilde{\Pi}_{2n} \, , 
\end{align} 
where $\tilde{\Pi}_{2n} \sim {\tilde{\cal R}}^n$ is a general endomorphism constructed with ambient curvatures. In reality, the endomorphism cannot be completely general:
for example, in $d=2$ the only possible modification is such that
$\tilde{\Pi}_{2} = c_1 {\tilde{R}}$, however ${\tilde{R}}=0$
in the ambient geometry, from which we conclude that the projection of $-\Box_{\tilde{g}}$ is the only available option.
The general rule is that we can only allow for modifications of the ambient operators that do not include ${\tilde{R}}$ and ${\tilde{R}_{AB}}$, given that the ambient space is Ricci-flat.

In the case $d=4$, we have that the only meaningful modification of the ambient Laplacian includes the square of the Riemann tensor, say, $\tilde{\Pi}_{4} = \xi_1 {\tilde{R}_{ABCD}\tilde{R}^{ABCD}}$ for constant $\xi_1$. We know already that ${\tilde{R}_{ABCD}\tilde{R}^{ABCD}}$ projects to $W_{\mu\nu\alpha\beta}W^{\mu\nu\alpha\beta}$ at $\rho=0$, so the replacement
$(-\Box_{\tilde{g}})^2 \to (-\Box_{\tilde{g}})^2 + \tilde{\Pi}_{4}$
corresponds to a replacement $(\Delta_4,Q_4) \to (\Delta_4,Q_4 +\xi_1 W^2)$, in which the $Q$-curvature is shifted by a Weyl covariant scalar.
Actually, this is true in general $d\geq 4$: a modification of the GJMS construction that involves an endomorphism constructed with the ambient Riemann tensor performs a replacement
\begin{align}
Q_{2n}
\rightarrow
\bar{Q}_{2n} 
=
Q_{2n} 
+
\sum_I \xi_I \, W_{2n,I}  \, ,
\end{align}
while leaving $\Delta_{2n}$ in $d=2n$ invariant. These are examples of \emph{fake} ambiguities because, even if we change the parametrization of the anomaly by adopting the new basis, this corresponds only to a linear redefinition of the charges of the anomaly and the final nonlocal action remains invariant because the change in the topological anomaly's charge is compensated by the opposite change in the $b$-anomalies. In other words, the integrated anomaly does not depend on the constant coefficients $\xi_I$ (recall that the integrated anomaly is defined up to conformally invariant terms).

Changes in the derivative part of the ambient Laplacians, instead, have the potential of introducing \emph{true} ambiguities, but only for $d\geq 6$. In $d=4$, we can convince ourselves that such ambiguities do not arise because the only feasible lower derivative terms would involve the ambient Ricci tensor and scalar, which actually vanish. In fact, a self-adjoint modification of the ambient Laplacian for $d=4$ spacetime dimensions would be
\begin{align}
\Box_{\tilde{g}}^2 \Phi \to \Box_{\tilde{g}}^2 \Phi + \tilde{\nabla}_A \Bigl(\bigl( c_1 \tilde{R}^{AB} + c_2 \tilde{R} \tilde{g}^{AB}\bigr)\tilde{\nabla}_B \Phi \Bigr)\,,
\end{align}
but, again, Ricci-flatness ensures that the differential operator is unchanged like in $d=2$. When playing this game, we constrain the leading highest derivative part of the operator, so that we ensure that its projection goes to $(-\partial^2)^2$ in the flat-space limit. Furthermore, when written in a manifestly self-adjoint form the operator does not have a term with three covariant derivatives. These two last points remain true in the general case.

In order to truly change the GJMS operators in a meaningful way, we need to be able to couple the ambient Riemann tensor to the derivative part. However, to do so, there must be enough derivatives
in the subleading parts of the operator, which have $2n-4$ or lower derivatives in $d=2n$, so that contractions with Riemann are possible (rather than with its traces). The first dimension in which this is possible is $d=6$, where we can consider a two-parameters family   
\begin{align}\label{eq:ambiguity_6d}
\tilde{O}_{6,d}
&=
(-\Box_{\tilde{g}})^3 
+
\alpha_1 \, \tilde{R}^{ABGR} \tilde{R}^{C}{}_{BGR} \, \tilde{\nabla}_{A} \tilde{\nabla}_{C}
+
\alpha_2 \, \tilde{R}^{ABGR}     \tilde{R}_{ABGR} \,  \tilde{ \Box} \, \\[5pt]  \nonumber
&+
 \alpha_1 \,  \tilde{R}^{ABGR} \left( \tilde{\nabla}_{A}  \tilde{R}^{C}{}_{BGR} \right) \tilde{\nabla}_{C}
+
2 \,  \alpha_2 \, \tilde{R}^{ABGR}   \left(     \tilde{\nabla}^{D}    \tilde{R}_{ABGR}    \right)   \tilde{\nabla}_{D}   
\, ,
\end{align} 
where $(\alpha_1,\alpha_2)$ are the two constant parameters and the operator is written in such a way that it is self-adjoint (without self-adjointness there would be a third parameter).

The operator $\tilde{O}_{6,d}$ is invariant under ambient diffeomorphisms, so, by construction, it defines a conformally covariant operator $O_{6,d}$ when projected
\begin{align}
O_{6,d} \varphi(x)
=
 t^{-\frac{6+d}{2}} \tilde{O}_{6,d} (t^{\frac{6-d}{2}}\varphi)
 |_{\rho=0} \, ,
\end{align} 
exactly like its $D_{6,d}$ counterpart.
Analogous generalizations are possible in higher dimensions with increasing number of parameters. Their construction is trivial, although lengthy in that we have to ensure that they are self-adjoint.

We remain interested to the case $n=3$ for simplicity, but what shown below generalizes to $n>3$.
The projection of $\tilde{O}_{6,d}$ still enjois the decomposition in derivative and nonderivative parts
\begin{align}
O_{6,d} 
\equiv
{\cal D}_{6,d}
+ 
\frac{1}{2} (d-6) \mathcal{P}_{6,d}
\, ,
\end{align}
where the pair $({\cal D}_{6,d},\mathcal{P}_{6,d})$ generalizes the pair $(\Delta_{6,d},Q_{6,d})$ sharing its transformation properties.
Given that $\tilde{O}_{6,d} |_{\alpha_i=0} = (-\tilde{\Box})^3 = \tilde{D}_{6,d}$, we can parametrize the operator through deviations of the $n=3$ GJMS operator in general $d$. Generalizing the general form that we used in the GJMS case, where we split derivative and nonderivative parts
\begin{align}
O_{6,d} 
=
{\cal D}_{6,d}
+ 
\frac{1}{2} (d-6) \mathcal{P}_{6,d}
=
\Delta_{6,d} 
+
\delta\Delta_{6,d}
+ 
\frac{1}{2} (d-6)( Q_{6,d} 
+ 
\delta Q_{6,d}) 
\, ,
\end{align} 
in which we ensure $\mathcal{P}_{6,d}|_{\alpha_i=0}=\Delta_{6,d}$ and $\mathcal{P}_{6,d}|_{\alpha_i=0}=Q_{6,d}$. We also have that, by construction, the deviations $\delta\Delta_{6,0}$ and $\delta Q_{6}$ are linear in the parameters $\alpha_i$.

Going finally to $d=6$, we define ${\cal D}_{6}\equiv {\cal D}_{6,6}$ and ${\cal P}_6 = {\cal P}_{6,6}$, where the latter is a generalization of the $Q_6$ curvature. We find 
\begin{align}  \nonumber
{\cal D}_{6} 
&=
\Box^3 
-
6 \mathcal{J} \Box^2
+
\left(
16 K^{\mu\alpha} \nabla_\mu \nabla_\alpha 
-
2 (\Box \mathcal{J})        
+  
\alpha_2 W^{\mu\nu\alpha\beta}W_{\mu\nu\alpha\beta}    
-    
32  K^{\mu\nu}K_{\mu\nu}    
+     
8\mathcal{J}^2     
+
4 (\nabla^\mu   \mathcal{J} ) \nabla_\mu      
\right) 
 \Box \\ \nonumber
&+
16 (\nabla^\rho K^{\mu\alpha} )\nabla_\rho  \nabla_\mu  \nabla_\alpha 
+
\left(
16 B^{\mu\rho} 
-
\alpha_1 W^{\mu}{}_{\nu\alpha\beta}W^{\rho\nu\alpha\beta}    
-
32 K_{\beta\nu}W^{\mu\beta\nu\rho}   
+
192 K^{\mu}{}_{\nu}K^{\nu\rho}    
-
 32 K^{\mu\rho} \mathcal{J}   
\right. \\   \nonumber
&       
+
16   (\nabla^\mu \nabla^\rho   \mathcal{J}  )  
\Big)  
\nabla_\mu  \nabla_\rho  
+
\Big(
4 (\nabla^\mu  \Box \mathcal{J} ) 
+
\frac{1}{2} (4\alpha_2 - \alpha1 ) W^{\rho\nu\alpha\beta} ( \nabla^\mu W_{\rho\nu\alpha\beta}  ) 
+
4 \alpha_1 W^{\rho\nu\alpha\mu}(\nabla_\rho K_{\nu\alpha})  \\   
&  
+
96 K^{\rho\nu}(\nabla_\nu K_{\rho}{}^{\mu} )
+
64 K^{\rho\mu}(\nabla_\rho  \mathcal{J}  )  
-
32 K^{\rho\nu}(\nabla^\mu K_{\rho\nu} ) 
-
16  \mathcal{J}  (\nabla^\mu  \mathcal{J}  ) 
 \Big)  
\nabla_\mu
 \, ,
\end{align}
which shows that the new differential operator is truly different for a general geometry.
The displacements from the standard GJMS construction are in $d=6$
\begin{align}
&\delta\Delta_{6,6}
=
\alpha_2 \left(
 W^{\mu\nu\alpha\beta} \, W_{\mu\nu\alpha\beta} \, \Box
 +
 2 W^{\mu\nu\alpha\beta} (\nabla^{\rho} W_{\mu\nu\alpha\beta}) \nabla_{\rho}
\right)
-
\alpha_1 \left(
 W^{\rho}{}_{\nu\alpha\beta} \, W^{\mu\nu\alpha\beta}   \nabla_{\rho}
\right.\\ \nonumber
&\hspace{2cm}\left.
+
 W^{\rho\nu\alpha\beta} (\nabla_{\rho}      W^{\mu}{}_{\nu\alpha\beta} )         
 -
4  W^{\mu\alpha\beta\rho}(  \nabla_{\rho}      K_{\alpha\beta})
 \right) \nabla_{\mu}  \\             
&\delta Q_{6,6} 
=
\alpha_2 \left(
 \mathcal{J} W_{\mu\nu\alpha\beta} \,  W^{\mu\nu\alpha\beta}
 -
 4K^{\mu\rho} W_{\mu}{}^{\nu\alpha\beta} \,  W_{\rho\nu\alpha\beta}
 - 
 8 C^{\alpha\beta\mu}C_{\alpha\beta\mu}
  - 
 8   W_{\nu\alpha\beta\rho}  \nabla^{\rho} \nabla^{\alpha} K^{\nu\beta} 
\right)\\
&\hspace{2cm}+
\alpha_1    \left(    3    C^{\alpha\beta\mu}C_{\alpha\beta\mu}   + 2      W_{\nu\alpha\beta\rho}  \nabla^{\rho} \nabla^{\alpha} K^{\nu\beta} 
\right) \nonumber
\, .
\end{align}   
Importantly, the new operator ${\cal D}_6$ and curvature ${\cal P}_6$ inherently share the same covariance properties of their counterparts of the GJMS construction. A similar generalization holds true in all dimensions $d=2n$, so there exist generalized $Q$-curvatures, which we denote as $\mathcal{P}_{2n}$ and satisfy integrability conditions, i.e., $[\delta_\sigma,\delta_{\sigma'}]\mathcal{P}_{2n}=0$. 
We see how the parametric families of generalizations give rise to ambiguities at the end of the analysis of the $d=6$ example given below.

\subsection{Nonlocal action in $d=6$ and ambiguities}\label{sect:d6}

In $d=6$ ($n=3$) the $b$-anomalies are the six-dimensional conformal scalars that we have already derived by using the ambient space approach. We parameterize these anomalies with the basis $W_{6,I}$ given in Sect.~\ref{sect:invariants}
as
$
\langle T  \rangle_b 
\equiv
\sum_{I=1}^{3}b_{I}      W_{6,I} 
$. The charges of the $b$-anomaly are known to be constrained by the Hofman-Maldacena bounds in the case of unitary CFTs \cite{Hofman:2008ar},
but, in principle, are general for arbitrary conformal theories.
The scheme-dependent trivial anomaly can be schematically represented as
\begin{align}\label{eq:trivial_anomaly_6d}
\langle T  \rangle_{a'}
\equiv\nabla_\mu J^\mu_6=\sum_{i=1}^{6} f_i B^i \,,
\end{align}
where the $B^i$ are six independent boundary terms of dimension six that satisfy the Wess-Zumino consistency conditions, as detailed in Appendix~\ref{sect:basis6d}.
The topological $a$-anomaly is given by the six-dimensional Euler density $E_6$.
Combining everything, we can express the vacuum expectation value of the trace of the energy-momentum tensor in the standard form as
\begin{align}\label{anomaly6d}
\langle T  \rangle 
=
\sum_{I=1}^{3}b_{I}      W_{6,I}
+
a E_6
+
\nabla_\mu J^\mu_6\,.
\end{align}

We now focus on the integration of this anomaly, beginning with the conformally flat case by leveraging the properties of the modified six-dimensional pondered Euler density. Afterward, we extend this construction to more general spacetimes.
The six-dimensional pondered Euler density is given by
\begin{align}\label{pondE6}
\tilde{E}_6
=
E_6
+
\nabla_\alpha J^\alpha_6 \,,
\end{align}
where $J^\alpha_6$, as derived by Anselmi \cite{Anselmi:1999uk}, takes the form
\begin{align}
J^\alpha_6
=
&-
\left(
\frac{408}{5} - 20 \zeta
\right)
R^\nu{}_{\mu}\nabla_\nu R^{\mu\alpha}
-
\left(
\frac{36}{25} - 2 \zeta
\right)
R^{\alpha\mu}\nabla_\mu R
+
\zeta\nabla^\alpha (R^2)\\ \nonumber
&+
\left(
\frac{144}{5} - 10 \zeta
\right)
\nabla^\alpha (R^{\mu\nu}R_{\mu\nu})
-
\frac{24}{5}\nabla^\alpha  \Box R \, .
\end{align}
Thus, the anomaly can be expressed in terms of $\tilde{E}_6$ on conformally flat spacetimes as
\begin{align}
\langle T  \rangle 
=
a_1 \tilde{E}_6
+
\sum_{i=1}^{6} {f'_1}_i B^i \, ,
\end{align}
where the $B_i$ are the same total derivative terms mentioned earlier, evaluated in the appropriate limit.
As discussed also in \cite{Anselmi:1999uk}, this expression can be readily integrated on conformally flat spacetimes, where $\sqrt{g}\langle T\rangle \propto \partial^6 \sigma$, by exploiting the conformal properties of the pondered Euler density. Through straightforward manipulations, one can show that a relation similar to Eq.~\eqref{pondE6} holds also between $Q_6$ and  $\tilde{E}_6$. Moreover, from their definitions, it is simple to see that both quantities transform identically under conformal transformations when the metric is decomposed as $g_{\mu\nu}=e^{2\sigma}\delta_{\mu\nu}$.

An important distinction from the $d=4$ case is that, in $d=6$, boundary terms involving the Riemann tensor can be constructed. However, such terms cannot appear in the original definition of Anselmi's $\tilde{E}_6$ because of the conformally flatness condition, which relates the Riemann tensor to the Ricci tensor. Thus, when this condition is relaxed, the quantity $\tilde{E}_6$ must be appropriately modified. The most natural generalization in such cases is $Q_6$, which, like $\tilde{E}_6$, transforms linearly and with a self-adjoint operator under conformal transformations, but in general spacetimes, not just conformally flat ones. As before, the ambient space framework tells us in advance that the relevant quantities can be $Q_6$ and the operator whith which it transforms, i.e., $\Delta_6$.

On general spacetimes the anomaly takes the form \eqref{anomaly6d}. We can explicitly verify that it is still possible to reexpress it using $Q_6$ instead of $E_6$ thanks to the relation 
\begin{align}\label{eq:relation_E6_Q6_general}
E_6
-6 \, Q_6
+
\sum_{i=1}^{7} f_i A_i
+
\sum_{I=1}^{3}  c_{I} W_{6,I}
=
0
\,,
\end{align}
where the $A^i$ are the seven possible boundary terms of dimension six reported in Appendix~\ref{sect:basis6d}.
This relation holds for the following constants
\begin{align}\label{eq:constants_Q6_general}
 & f_1 =\frac{3}{5} \, , 
 && f_2 =-\frac{48}{25}\, ,
 &&  f_3=3\, ,
 && f_4=12\,
 &&  f_5=-6\, ,
 \\
 & f_6= -12\, , 
 && f_7=-3\,  
 &&c_{1}=-3 \, ,
 && c_{2}= 12 \, ,
  && c_{3}= 1  \, . 
  \nonumber
\end{align}
As previously discussed, the total derivative in the linear relation between the $Q$-curvature and the Euler density is required to satisfy the Wess-Zumino consistency conditions in each even dimension. Consequently, it must be possible to express $\sum_{i=1}^{7} f_i A_i=\sum_{i=1}^{6} k_i B_i$. We explicitly verify this fact in the $d=6$ case in Appendix~\ref{sect:basis6d}.

Thus, the anomaly can now be conveniently expressed as
\begin{align}\label{anomaly6d-Q6}
\langle T  \rangle 
&=
\sum_{I=1}^{3}\tilde{b}_{I}      W_{6,I}
+
\tilde{a} Q_6
+
\sum_{i=1}^{6} \tilde{f}_i B^i \,,
\end{align}
where $\tilde{a}=6 a$, and the constants $\tilde{b}_{I}$ and $\tilde{f}_i$ are obtained by constant shifts according to Eqs.~\eqref{eq:constants_Q6_general} and the results of Appendix~\ref{sect:basis6d}. Once again, using the known covariance properties of $Q_6$, the anomaly becomes easily integrable in this form.

Focussing on the first four terms, which give the nonlocal part of the integrated $\Gamma$, we find using the Riegert gauge
\begin{align}\label{eq:6d_NL_action_Q_curvature}
\Gamma_{\rm nl}[g]
=
-  \int \dd^6x\, \sqrt{g} \left(
\sum_{I=1}^{3}\tilde{b}_{I}      W_{6,I}  +\frac{\tilde{a}}{2} Q_6  
\right)
\frac{1}{\Delta_6} Q_6 \,,
\end{align}
where $1/\Delta_6$ is the Green function of $\Delta_6$ \cite{Osborn:2015rna}.

Now, our main claim is that the integrated nonlocal action is ambiguous because of our arbitrary choice of the pair $(\Delta_6,Q_6)$, and this ambiguity is a true one. To confirm this, we use ${\cal D}_6$ and $\mathcal{P}_6$ to integrate the anomaly in place of $\Delta_6$ and $Q_6$. For instance, Eq.~\eqref{conf_transf_Q} remains unchanged in form if we simultaneously substitute $\Delta_6 \to O_6$ and $Q_6 \to \mathcal{P}_6$.
The change affects the trivial anomaly, so it changes a local term that depends on the scheme in $\Gamma$.
It is simple to see that the anomaly can be reparametrized also in terms of $\mathcal{P}_6$ instead of $E_6$ since
\begin{align}
E_6
- 6\, \mathcal{P}_6
+
\sum_{i=1}^{7} f_i A_i
+
\sum_{I=1}^{3}  c_{I} W_{6,I}
=
0
\end{align}
is satisfied for the the following choice of the constants
\begin{align}\label{eq:constants_P6_general}
 & f_1 =\frac{3}{5} \, , 
 && f_2 =-\frac{3}{200}\left(128-5\alpha_1+20\alpha_2 \right)\, ,
 &&  f_3=\frac{3}{8}\left( 8-\alpha_1   \right) \, ,
 \\ \nonumber
 & f_4=\frac{3}{4}\left( 16 -\alpha_1 + 4\alpha_2   \right) \,,
 &&  f_5=-\frac{3}{4}\left( 8 -\alpha_1  \right)\, ,
 && f_6= -3\left( 4 - \alpha_1  \right)\, , 
  \\ \nonumber
 & f_7=-3\left( 1+ \alpha_1  \right)\, , 
 &&c_{1}=-3+\alpha_2 \, ,
 \\ \nonumber
 & c_{2}= 4(3+\alpha_2) \, ,
 && c_{3}= 1+\alpha_2  \, . 
  \nonumber
\end{align}
As before, the anomaly can  be expressed as $\langle T  \rangle 
=
\sum_{I=1}^{3} \tilde{b}_{I}      W_{6,I}
+
\tilde{a} \mathcal{P}_6
+
\sum_{i=1}^{6} \tilde{f}_i B^i$, see also Appendix~\ref{sect:basis6d}. However, the charges now differ from
the previous choice in that they depend on the parameters $\alpha_i$.
Following the usual steps, we find the nonlocal part of the anomalous action
\begin{align}\label{eq:6d_NL_action_P}
\Gamma_{\rm nl}[g]
=
-  \int \dd^6x\, \sqrt{g} \left(
\sum_{I=1}^{3} \tilde{b}_{I}      W_{6,I}  +\frac{\tilde{a}}{2} \mathcal{P}_6  
\right)
\frac{1}{{\cal D}_6} \mathcal{P}_6 \,.
\end{align}
Although Eqs.~\eqref{eq:6d_NL_action_P} and \eqref{eq:6d_NL_action_Q_curvature} both yield essentially the same anomaly (up to trivial anomalies), they are two genuinely different actions, in that the former exhibits a nontrivial dependence on the parameters $\alpha_i$ through ${\cal D}_6$ and $\mathcal{P}_6$.

Interestingly, a similar result was derived through significantly different approachs in Refs.~\cite{Hamada:2000me} and \cite[PRD version]{Ferreira:2017wqz}.
Our operator ${\cal D}_6$ can be shown to correspond to those obtained in Refs.~\cite{Hamada:2000me,Ferreira:2017wqz} by a trivial reparametrization of the constants $\alpha_i$. However, from the perspective of the ambient space, these parametric ambiguities have a well-defined geometric origin, which are rooted in the possibility to generalize the GJMS hierarchy.
This implies, for example, that such ambiguities will manifest in a structurally similar manner across higher dimensions, analog to the $d=6$ case.

A relevant implication of these ambiguities is that certain physical quantities may depend on these parameters. For instance, the part of the full energy-momentum tensor derived from Eq.~\eqref{eq:6d_NL_action_P} will explicitly depend on the $\alpha_{i}$s. Beyond their clear geometrical origin, it would be of significant interest to determine whether these parameters have a physical interpretation -- for example in the contexts of black hole thermodynamics or inflation -- or if they can and should be reabsorbed in a conformal action. In a recent work \cite{Paci:2024ohq}, we have shown that similar ambiguities of integrated nonlocal actions cancel in the computation of the Wald entropy of a black hole, so it is possible that at least some observables are not affected by these parameters.

\section{Conclusions and future perspectives}\label{sect:conclusions}

In this paper we have discussed how the ambient space construction by Fefferman and Graham can be used to construct the possible curvature monomials that satisfy the integrability conditions for the trace anomaly of Weyl invariance, i.e., the $a$- and $b$-anomalies. 
The important role that the ambient space has in providing integrable terms for the anomaly has been already established in the past, however, in our analysis, we clarify how the contribution of the topological $a$-anomaly in the ambient approach, here identified with the $Q$-curvature, always comes ``paired'' with a conformally covariant GJMS operator $\Delta_d$, which is a fundamental ingredient for its integration. In fact, we can write the nonlocal action, that generalizes Polyakov's and Riegert's, which integrates the anomaly quite compactly using all these ingredients.

Pragmatically, our analysis highlights the ambient geometrical origin of the nonlocal action and shows that there are potential ambiguities in the integration that, for the case in which the only nondynamical source is the metric, emerge in spacetime dimensions higher than six. These ambiguities show that the integrated nonlocal action is less ``predictive'' in higher dimensions, in the sense that the trace anomaly cannot determine the nonlocal action completely by itself. The higher is the dimension $d$, the higher is the number of undetermined parameters of a nonlocal action integrating the anomaly. In a sense, our conclusions are underwhelming, in that the integration of the anomaly seems to work very well in the case $d=4$ which, as it happens, is probably the most physically interesting case. We should note that the case $d=6$ (for which the first ambiguities arise) is often regarded as quite interesting, but, probably, more for theoretical reasons than phenomenological ones.

There is an elephant in the room, however, and this conclusion gives us the opportunity to discuss it. While being very potent for the computation of the natural $a$- and $b$-anomalies, strictly speaking the ambient space does not really exist in even dimensions, at least in the way in which the Ricci-flatness condition is solved as a perturbative Taylor expansion in the ambient coordinate $\rho$. Geometrically, this is seen through the appearance of special ``obstructions'' which are curvature tensors that essentially prevent the lift of a flat-space CFT to full curved space Weyl invariance. In other words, the integrable terms of the anomaly are extracted from a Taylor expansion in general $d$ that is actually nonanalytic for an arbitrary geometry in $d=2n$. Even though the integrability properties of the $a$- and $b$-anomalies candidates can be verified manifestly and hold correctly for $d=2n$, despite the obstructions, we are always secretly circumventing the obstructions by working in arbitrary $d$, taking the even dimension limit $d=2n$ at the end to determine the actual anomaly's terms. Whether this operation of taking the limit actually commutes with the integration of the anomaly to a nonlocal action at any order of a curvature expansion is, in our humble opinion and in the opinion of Ref.~\cite{Duff:1993wm}, up for debate.

If we are willing to always and only work in $d=2n$, the original Fefferman and Graham's expansion should be at least replaced by one that would include logarithms of the coordinate $\rho$ as shown in Ref.~\cite{graham-hirachi}, making it a trans-serie, or, alternatively, the Ricci-flatness condition should be solved by other means. Given that the Taylor expansion seems sufficient to produce the integrable terms and that the nonlocal actions seem to work in the physical dimension $d=4$, this problem seems more scholastic than pragmatical. However, the logarithmic terms are used in the AdS/CFT formulation of the ambient approach, which we have touched in our discussion, precisely as a mean to produce from the bulk action the integrable terms.
These considerations may have a role in understanding the actual physical value of the nonlocal actions.\footnote{%
By this we do not at all mean that the nonlocal actions do not have physical value already, in fact they have been applied very successfully to several physical contexts: black holes, inflation, string theory etc., see for example Ref.~\cite{Duff:1993wm} and references therein.
}

It has been shown in Ref.~\cite{Coriano:2022jkn} through an analysis of the $\langle TTJJ\rangle$ correlator -- where $J$ represents a conserved Abelian current -- that in $d=4$ the nonlocal actions are unable to reproduce the anomalous Ward identities derived from perturbation theory for free field theories around flat space.
In essence, there is a discrepancy between two approaches: on the one hand we have a geometrical framework to generate nonlocal actions that should, in principle, satisfy all constraints
of the $n$-point correlators of the energy-momentum operator determined by $\langle T \rangle \neq 0$ and diffeomorphism invariance; on the other hand, we have an explicit computation in flat (momentum) space of the same correlators as constrained by CFT. 
While the two approaches are consistent at the level of three-point functions \cite{Coriano:2017mux}, they do not match for four-point functions. Specifically, using the names introduced in Sect.~\ref{sect:d4}, the Riegert-Fradkin-Tseytlin gauge choice introduces problematic double poles, while the Fradkin-Vilkovisky gauge, despite avoiding this issue, still does not agree with perturbative results. Ad hoc modifications of the nonlocal action have been proposed in Ref.~\cite{Coriano:2022jkn} that reproduce the CFT results, but their Weyl geometrical origin is unclear to our knowledge.
It is a legitimate conjecture, which resonates with the early works on the anomaly by Duff and collaborators (see Ref.~\cite{Duff:1993wm} for a historical perspective), that
the limit $d\to 2n$ may still be hiding something which, arguably, is caused by the obstructions, and that this something may explain the differences observed in Ref.~\cite{Coriano:2022jkn}. We certainly hope that this problem is further investigated in the future.

\smallskip

\paragraph*{Acknowledgments.}

We are grateful to Davide~Pittet for collaboration on topics related to this project during the development of his MSc thesis on ambient space and Ricci-gauging at the University of Pisa \cite{pittet}. OZ is also grateful to Andreas~Stergiou for early clarifications on the role of $Q$-curvatures that instigated the beginning of this work.
We are also grateful to Claudio Corian\`o for important discussions that motivated us to conclude this project and to Riccardo~Martini for valuable discussions. Finally we thank Danilo D\'iaz V\'azquez for a comment that sparked an improvement of the analysis of the trivial anomaly in the draft.

\appendix

\section{Conformal tensors}\label{sect:conformal-tensors}

The expansion tensors of Sect.~\ref{sect:ambient-main} are best expressed in terms of conformally covariant tensors and, in general, tensors with special conformal properties. In fact, the ambient space formalism can be seen as a geometrical construction that allows to find and classify such tensors.
We follow the conventions of Ref.~\cite{Osborn:2015rna} for the various definitions. Of paramount importance are the Schouten tensor and its trace
\begin{equation}
\begin{split}
 K_{\mu\nu} = \frac{1}{d-2}\Bigl(R_{\mu\nu}-\frac{1}{2(d-1)} R g_{\mu\nu}\Bigr)\,,
 \qquad {\cal J} = g^{\mu\nu} K_{\mu\nu} = \frac{1}{2(d-1)} R\,.
\end{split}
\end{equation}
They enjoy the Weyl transformations
$\delta_\sigma K_{\mu\nu}=-\nabla_\mu \partial_\nu\sigma$
and $\delta_\sigma {\cal J} = -2{\cal J} \sigma -\nabla^2 \sigma$.
The Weyl tensor is defined
\begin{equation}
\begin{split}
 W_{\mu\nu \lambda\theta} = R_{\mu\nu\lambda\theta} -g_{\mu\lambda} K_{\nu\theta} +g_{\nu\lambda} K_{\mu\theta} - g_{\nu\theta}K_{\mu\lambda} + g_{\mu\theta} K_{\nu\lambda} \,,
\end{split}
\end{equation}
and it is invariant under Weyl transformations when one index is raised, $\delta_\sigma W_{\mu\nu}{}^\lambda{}_\theta=0$.
The Cotton and Bach tensors are defined
\begin{equation}
\begin{split}
 &C_{\mu\nu\lambda}=\nabla_\lambda K_{\mu\nu}-\nabla_\nu K_{\mu\lambda}
 \,, \qquad
  B_{\mu\nu} = \nabla^\lambda C_{\mu\nu\lambda} +K^{\lambda\theta} W_{\lambda\mu\theta\nu}\,.
\end{split}
\end{equation}
They enjoy the Weyl transformations
$\delta_\sigma C_{\mu\nu\lambda}= -\nabla^\theta \sigma W_{\theta\mu\nu\lambda}$ and $\delta_\sigma B_{\mu\nu} = -2 \sigma B_{\mu\nu}+(d-4)\nabla^\theta \sigma(C_{\mu\nu\theta}+C_{\nu\mu\theta})$.
Since the Weyl tensor is automatically zero in $d=3$, because the Riemann tensor is completely determined by the Ricci tensor, we have that the Cotton tensor is conformally invariant in $d=3$ and in a sense plays the role of the Weyl tensor in $d=3$. The Cotton tensor has also mixed-symmetry (hook-symmetry) properties
\begin{equation}
 C_{\mu\nu\lambda}= - C_{\mu\lambda\nu} \,,        \qquad      C_{\mu\nu\lambda} + C_{\nu\lambda\mu} +  C_{\lambda\mu\nu} = 0 \,,        \qquad      C^{\mu}{}_{\mu\lambda} = 0   \,,      \qquad      \nabla^\mu C_{\mu\nu\lambda} = 0 \,.
\end{equation}
Furthermore, the Bach tensor is invariant (more precisely it is Weyl-covariant with weight $-2$) in $d=4$. The Bach tensor is symmetric, traceless and divergenceless in any $d$ (i.e., $B_\mu{}^\mu=\nabla^\mu B_{\mu\nu}=B_{[\mu\nu]}=0$). In the main text it is clarified that the invariance of the Bach tensor in $d=4$ comes as a result of geometrical obstructions to the lift of flat space's conformal invariance to curved space's Weyl invariance \cite{Karananas:2015ioa}. Further tensors generalizing the Bach tensor in any even dimension higher than four can be constructed, but they can all be expressed in terms of the four basic tensors defined in this appendix.

\section{Basis for dimension six scalars and trivial anomalies in $d=6$}\label{sect:basis6d}

It has been established that there are $17$ independent invariants $I_i$ of order six \cite{Fulling:1992vm}.
A convenient basis for these invariants is 
\begin{align}\label{tot_invariants}
\setlength{\jot}{10pt}
 & I_1 =\Box^2 R \, , 
 && I_2 =R\Box R\, ,
 && I_3=R^{\mu\nu}\Box R_{\mu\nu}\, ,
 && I_4=   R^{\mu\nu}\nabla_{\mu}\nabla_{\nu} R   \,,
 \nonumber\\
 & I_5=R^{\mu\nu\alpha\beta} \Box R_{\mu\nu\alpha\beta}\, ,
 && I_6= \nabla^{\mu}R\nabla_{\mu}R    \, , 
 && I_7=\nabla^{\mu}R^{\nu\rho}\nabla_{\mu}R_{\nu\rho} \, ,
 && I_8= \nabla^{\mu}R^{\nu\rho}\nabla_{\rho}R_{\mu\nu} \,  ,
 \nonumber\\
 & I_9=\nabla^{\mu}R^{\nu\rho\gamma\sigma}\nabla_{\mu}R_{\nu\rho\gamma\sigma} \, ,
 && I_{10}= R^3 \, ,
 && I_{11}=R R^{\mu\nu} R_{\mu\nu}\, ,  
 && I_{12}= R^{\mu\nu}R_{\nu\rho}R^{\rho}_{\mu}\, , 
   \nonumber\\
 & I_{13}= R^{\mu\nu}R^{\gamma\rho} R_{\mu\gamma\nu\rho}     \, ,
 && I_{14}= R R^{\mu\nu\gamma\rho} R_{\mu\nu\gamma\rho} \,  ,
 && I_{15}= R^{\mu\nu}R_{\mu}{}^{\rho\gamma\sigma}R_{\nu\rho\gamma\sigma}  \, ,
 &&I_{16}= R^{\alpha\beta\gamma\rho}R_{\gamma\rho\mu\nu}R^{\mu\nu}{}_{\alpha\beta} \, ,
   \nonumber\\
 & I_{17}= R^{\alpha\beta\gamma\rho}R_{\alpha\mu\gamma\nu}R^{\mu}{}_{\beta}{}^{\nu}{}_{\rho}  \, .
\end{align}      
However, upon considering integration by parts, it becomes evident that only $10$ of these invariants are independent. To illustrate this, note that instead of $I_5$, one could equivalently use the scalar $I'_5=R_{\mu\nu\rho\gamma}\nabla^{\gamma}\nabla^{\nu} R^{\mu\rho}$, due to the identity given in Eq.~\eqref{eq:RiemBoxRiem}.
By applying the Bianchi identities and performing integrations by parts, we can construct a more compact basis as follows
\begin{align}\label{eq:compact_basis_6d}
\setlength{\jot}{10pt}
 & I_1 =R\Box R\, ,
 && I_2=R^{\mu\nu}\Box R_{\mu\nu}\, ,
 && I_{3}= R^3 \, ,
 &&  I_{4}=R R^{\mu\nu} R_{\mu\nu}\, ,  
 \nonumber\\
 &  I_{5}= R^{\mu\nu}R_{\nu\rho}R^{\rho}_{\mu}\, , 
 && I_{6}= R^{\mu\nu}R^{\gamma\rho} R_{\mu\gamma\nu\rho}     \, ,
 && I_{7}= R R^{\mu\nu\gamma\rho} R_{\mu\nu\gamma\rho} \,  ,
 && I_{8}= R^{\mu\nu}R_{\mu}{}^{\rho\gamma\sigma}R_{\nu\rho\gamma\sigma}  \, ,
  \nonumber\\
 &I_{9}= R^{\alpha\beta\gamma\rho}R_{\gamma\rho\mu\nu}R^{\mu\nu}{}_{\alpha\beta} \, ,
 && I_{10}= R^{\alpha\beta\gamma\rho}R_{\alpha\mu\gamma\nu}R^{\mu}{}_{\beta}{}^{\nu}{}_{\rho}  \, .
\end{align}    
As we could expect,
we can construct $7$ total derivatives terms, which can be systematically obtained by considering the dimension $5$ vectors. A convenient choice for this basis is
\begin{align}\label{eq:boundary_terms}
\setlength{\jot}{10pt}
 & A_1 =\Box^2 R \, , 
 && A_2 =\nabla_{\mu}\left(    R  \nabla^\mu R  \right)\, ,
 && A_3=\nabla_{\mu}\left(    R^{\mu\nu}\nabla_\nu R    \right)\, ,
 \nonumber\\
 & A_4=  \nabla_{\mu}\left( R^{\rho\gamma}  \nabla^\mu   R_{\rho\gamma}   \right) \, ,
 && A_5=\nabla_{\mu}\left(   R^{\rho\gamma}  \nabla_\gamma   R^{\mu}_{\rho}    \right)\, ,
 && A_6=  \nabla_{\mu}\left(   \nabla^{\nu} R^{\rho\gamma} R^{\mu}{}_{\rho\gamma\nu}  \right)  \, , 
  \nonumber\\
 & A_7= \nabla_{\mu}\left(  R^{\alpha\beta\gamma\rho} \nabla^\mu R_{\alpha\beta\gamma\rho}   \right)      \, .
\end{align}
However, not all of these boundary terms individually satisfy the Wess-Zumino consistency conditions. As previously noted in Ref.~\cite{Bastianelli:2000rs}, the basis of integrable trivial anomalies consists of six independent terms. This result can be readily derived also using the approach outlined in Ref.~\cite{Bonora:1983ff,Bonora:1985cq}, where the field $\sigma$ is treated as a Grassmannian function, and the nilpotent BRST differential for Weyl transformations $s$ is introduced. The BRST differential acts on the metric and the Riemann tensor as follows
\begin{align}\label{eq:BRST_transformation_riemann}
s\, g_{\mu\nu}=2\sigma g_{\mu\nu}\, , \qquad
s\, R^{\mu}{}_{\nu\rho\gamma}
=
g_{\rho\nu}\nabla^{\mu}\nabla_{\gamma}\sigma -g_{\gamma\nu}\nabla^{\mu}\nabla_{\nu}\sigma
+
\delta^{\mu}{}_{\gamma}\nabla_{\nu}\nabla_{\rho}\sigma - \delta^{\mu}{}_{\rho}\nabla_{\nu}\nabla_{\gamma}\sigma \, .
\end{align}
Following Ref.~\cite{Bonora:1983ff,Bonora:1985cq}, the consistency conditions for the boundary terms can then be written as
\begin{align}\label{eq:CC_BT}
s \sum_{i=1}^{7}\int \dd^{6} x \sqrt{g} \sigma f_i A_i
=
0 \, ,
\end{align}
where the $A_i$ are those of Eq.~\eqref{eq:boundary_terms}. After some tedious but straightforward computations, from Eq.~\eqref{eq:CC_BT} we obtain
\begin{align}
(10 f_3 - 4f_4 -3f_5 + f_6 - 4f_7)\,\sigma \left( \nabla_\mu \nabla_\nu \Box \sigma \right) R^{\mu\nu}
-
\frac{1}{2}(10 f_3 - 4f_4 -3f_5 + f_6 - 4f_7)\,\sigma \left( \Box^2 \sigma \right) R
=
0 \, .
\end{align}
Thus, we deduce the following relation
\begin{align} \label{eq:constraint_constants_BT}
f_6= 4 (f_4 + f_7) + 3 f_5 - 10 f_3 \, ,
\end{align}
which agrees with the results of Ref.~\cite{Bastianelli:2000rs}.
 According to Eq.~\eqref{eq:constraint_constants_BT}, the basis of integrable trivial anomalies can be expressed as
\begin{align}\label{eq:integrable_boundary_terms}
\setlength{\jot}{10pt}
 & B_1 = A_1\, , 
 && B_2 =A_2\, ,
 && B_3= A_3 - 10 A_6  \, ,
 \nonumber\\
 & B_4=  A_4 + 4A_6 \, ,
 && B_5= A_5 + A_6\, ,
 && B_6=A_7 + 4 A_6  \, .
\end{align}
Furthermore, it is straightforward to verify that the specific combinations of boundary terms appearing in Eq.~\eqref{eq:relation_E6_Q6_general} and Eq.~\eqref{eq:constants_P6_general} satisfy Eq.~\eqref{eq:constraint_constants_BT}, and thus they can be written in the basis defined in Eq.~\eqref{eq:integrable_boundary_terms}.

As a cross-check, we present a local action in the basis of Eq. \eqref{eq:compact_basis_6d}  that reproduces the total derivatives in Eq.~\eqref{eq:relation_E6_Q6_general} of the main text. A direct computation confirms that
\begin{align} 
 \Gamma^{Q_6}_{\rm an,loc}
 =
-
 \int \dd^6x\, \sqrt{g} \left(
 \frac{1}{10} I_2
 -
  \frac{47}{45} I_5
 -
   \frac{34}{75} I_6
 -
   \frac{11}{30} I_7
 +
   \frac{452}{75} I_8
  -
   \frac{1483}{450} I_9
 \right)
\end{align} 
is such that
\begin{align} 
 2 g_{\mu\nu} \frac{\delta}{\delta g_{\mu\nu}}  \Gamma^{Q_6}_{\rm an,loc} = \sqrt{g} \sum_{i=1}^{7} f_i A_i \, ,
\end{align} 
where the constants $f_i$ are those of Eq.~\eqref{eq:constants_Q6_general}.

For completeness, we conclude this appendix by presenting the Euler density in six dimensions, expressed in terms of conformal tensors
\begin{align}
E_6\nonumber
&=
\frac{1}{8} R_{\alpha\beta\tau\xi} R_{\mu\nu\gamma\omega} R_{\rho\sigma\lambda\zeta} \epsilon^{\alpha\beta\mu\nu\rho\sigma}\epsilon^{\tau\xi\gamma\omega\lambda\zeta}\\
&=48 \mathcal{J}^3 
+
96 K^{\mu}{}_{\nu} \, K^{\rho}{}_{\mu} \, K^{\nu}{}_{\rho} 
-
144 K^{\mu\nu} \, K_{\mu\nu} \mathcal{J}
+
48  K^{\mu\nu} \, K^{\rho\gamma} \, W_{\mu\rho\nu\gamma}
+
6 \mathcal{J}\, W^{\mu\rho\nu\gamma} \, W_{\mu\rho\nu\gamma} \\
& \quad-
24 K_{\alpha\beta} \, W^{\alpha\rho\nu\gamma} \, W^{\beta}{}_{\rho\nu\gamma}
-
8 W^{\alpha\beta\gamma\rho}\,W_{\alpha\mu\gamma\nu}\,W^{\mu}{}_{\beta}{}^{\nu}{}_{\rho}
+
2 W^{\alpha\beta\gamma\rho}\,W_{\gamma\rho\mu\nu}\,W^{\mu\nu}{}_{\alpha\beta}\, . \nonumber
\end{align}
The complete basis \eqref{tot_invariants} can also be used to obtain the three $b$-anomalies in $d=6$. The procedure involves writing them as ambient tensors with the replacement $R_{\cdots} \to \tilde{R}_{\cdots}$, using then Ricci-flatness we are left with four tensors that depend only on the ambient Riemann $\tilde{R}_{ABCD}$, but one of them can be further simplified using the Bianchi identities as explained in the main text, leading to the basis $\tilde{W}_{6,I}$ for $I=1,2,3$ used in the main text.

\section{Consistency conditions for the Euler density in arbitrary even dimensions}\label{sect:CC_Ed}

In this appendix, we verify directly that the Euler density satisfies the Wess-Zumino consistency conditions in arbitrary even dimensions. For a broader perspective, see Ref.~\cite{Boulanger:2007st}, in which the general solutions of the so-called Stora-Zumino chain of descendant equations for the trace anomaly are derived. Here, we less ambitiously focus on simply checking that the Euler density satisfies the Wess-Zumino consistency conditions in any even $d$. To streamline the computation, we adopt again the approach from Ref.~\cite{Bonora:1983ff,Bonora:1985cq} outlined in Appendix \ref{sect:basis6d}. 

We recall that the Euler density in $d=2n$ dimensions is defined as
\begin{align}
  E_{2n}(g) = \frac{1}{2^n} R_{\mu_1\mu_2\nu_1\nu_2} \cdots R_{\mu_{2n-1}\mu_{2n}\nu_{2n-1}\nu_{2n}} \epsilon^{\mu_1\cdots\mu_{2n}}\epsilon^{\nu_1\cdots\nu_{2n}} 
  \,,
\end{align}
and our objective is to check that $s\int \dd^{2n} x \sqrt{g} \sigma E_{2n}=0$, which then implies integrability. The key result that we rely on is that
\begin{align}\label{eq:basic_relation}
\int  \dd^{2n} x \sqrt{g} N^{\mu\nu} \sigma \nabla_\mu \nabla_\nu \sigma
=
\int  \dd^{2n} x \sqrt{g} N^{(\mu\nu)} \sigma \nabla_\mu \nabla_\nu \sigma
= - \int  \dd^{2n} x \sqrt{g}  N^{(\mu\nu)}  \nabla_\mu\sigma  \nabla_\nu \sigma
=0 \, ,
\end{align}
which is true if $\nabla_\mu N^{\mu\nu} =0$, thanks to the now Grassmannian nature of $\sigma$.
With this result in mind, we compute
\begin{align}
s\int \dd^{2n} x \sqrt{g} \sigma E_{2n}
 &=\frac{1}{2^n} \int \dd^{2n} x \sqrt{g} \sigma \epsilon_{\mu_1}{}^{\mu_2}{}_{\cdots\mu_{2n}}\epsilon^{\nu_1\cdots\nu_{2n}}
 \sum_{j=1}^n \left[ 
  s R^{\mu_{\mu_{2j-1}}}{}_{\mu_{2j}\nu_{2j-1}\nu_{2j}}         \prod_{i\neq j=1}^{n-1} R^{\mu_{\mu_{2i-1}}}{}_{\mu_{2i}\nu_{2i-1}\nu_{2i}}     
  \right] \\ \nonumber
 &=  \frac{1}{2^{n-2}}  \int \dd^{2n} x \sqrt{g}  \epsilon^{\mu_1\cdots\mu_{2n}}\epsilon^{\nu_1\cdots\nu_{2n}} 
 \sum_{j=1}^n  g_{\mu_{2j}\nu_{2j-1}}   \left[ 
 \sigma  \nabla_{\mu_{2j-1}}   \nabla_{\nu_{2j}}  \sigma      \prod_{i\neq j=1}^{n-1} R_{\mu_{\mu_{2i-1}\mu_{2i}\nu_{2i-1}\nu_{2i}}}
 \right] \, .
\end{align}
In the last line, we used Eq.~\eqref{eq:BRST_transformation_riemann} and the symmetry properties of the Riemann tensor. Note that the integrand is symmetric under the exchange $(\mu_{2j-1} \leftrightarrow \nu_{2j})$ similarly to Eq.~\eqref{eq:basic_relation}.  For brevity, we omit explicitly denoting anti-symmetrization brackets induced by the $\epsilon$-products, but this should be understood. Next, we integrate by parts and find
\begin{align}
 s\int \dd^{2n} x \sqrt{g} \sigma E_{2n} &=
 -  \frac{1}{2^{n-2}}  \int \dd^{2n} x \sqrt{g} \epsilon^{\mu_1\cdots\mu_{2n}}\epsilon^{\nu_1\cdots\nu_{2n}} \,\times
 \nonumber\\&
 \times \sum_{j=1}^n g_{\mu_{2j}\nu_{2j-1}}  \left[
     \nabla_{\nu_{2j}} \sigma  \nabla_{\mu_{2j-1}}   \sigma      \prod_{i\neq j=1}^{n-1} R^{\mu_{\mu_{2i-1}}}{}_{\mu_{2i}\nu_{2i-1}\nu_{2i}}     
    +   \sigma  \nabla^{\mu_{2j-1}}   \sigma      \nabla^{\nu_{2j}}  \prod_{i\neq j=1}^{n-1}  R_{\mu_{\mu_{2i-1}\mu_{2i}\nu_{2i-1}\nu_{2i}}}
  \right] \, .
\end{align}
Here, the terms involving derivatives of the Riemann tensor vanish due to the differential Bianchi identity, enforced by the antisymmetry of $\epsilon^{\nu_1\cdots\nu_{2n}} $, which ensures  $\nabla_\mu N^{\mu\nu} =0$ when comparing with Eq.~\eqref{eq:basic_relation}. 
Additionally, the contribution involving $ \nabla_{\nu_{2j}} \sigma  \nabla_{\mu_{2j-1}} \sigma $ also vanishes because $\sigma$ is anticommuting, leading to a contraction of a symmetric tensor with an antisymmetric tensor.
We conclude that $s\int \dd^{2n} x \sqrt{g} \sigma E_{2n}=0$.

\section{Conformal anomaly and heat kernel coefficients}\label{sect:KH_anomaly}

In this appendix we rederive the equation that relates the anomalous trace of the energy-momentum tensor $\langle T \rangle $ to the coefficients of the heat kernel expansion of some operator in curved space. The approach involves comparing the variation of the effective action $\Gamma[g]$ under a Weyl transformation, when computed with two distinct strategies.
This allows us to present the basic set of assumptions that underline a general computation of the anomaly, at least in the ``unbroken'' phase of conformal invariance \cite{Schwimmer:2023nzk}, and, consequently, how such assumptions could potentially be challenged (see the discussion of Ref.~\cite[Sect.~7]{Duff:1993wm} and the manipulations of Ref.~\cite{Mukhanov:2007zz}).

On the one hand, consider that the variation of $\Gamma[g]$ with respect to the metric $g_{\mu\nu}$ sources $\langle T^{\mu\nu} \rangle$, so we have 
\begin{align}\label{1st_expr_Gamma}
\delta_\sigma \Gamma
=
\frac{1}{2} \int \dd^d x \, \sqrt{g} \,
\langle T^{\mu\nu} \rangle  \, \delta_\sigma g_{\mu\nu}
=
\int \dd^d x \, \sqrt{g} \,
\langle T  \rangle \, \sigma(x) \,,
\end{align}
where we used $ \delta_\sigma g_{\mu\nu} = 2 \sigma  g_{\mu\nu}$.  On the other hand, the effective action can be expressed in terms of the zeta function $\zeta (r) = \sum_n \lambda_n^{-r}$ of some Hessian operator $\hat{O}(g)$, as
\begin{align}
\Gamma[g+\delta g]
=
-\frac{1}{2}
\frac{\dd}{\dd r}  \zeta_{\hat{O}} (g+\delta g ; r)  |_{r=0}
\,,
\end{align}
where $ \zeta_{\hat{O}} (g;r)$ is computed from the eigenvalues $\lambda_n$ of the differential operator $\hat{O}(g)$. The dependence on $g$ is to stress that it is constructed with the covariant derivatives associated with the metric $g$. In specific applications the Hessian is the second functional derivative of the action of some fields $\phi^i$ that we are integrating-out.

The zeta function can be represented in terms of an integral over a fictituous time $\tau$
\begin{align}
\zeta_{\hat{O}} (g;r)
&=
{\rm Tr} \big\{    {\hat{O}(g)^{-r}} \big\}
=
 \int \dd^d x \, \sqrt{g} 
 \langle x | \hat{O}(g)^{-r} | x \rangle\\ \nonumber
&\equiv
 \int \dd^d x \, \sqrt{g} 
\,  \zeta_{\hat{O}} (g;r|x,x)   
=
\frac{1}{\Gamma(r)} \int_{0}^{\infty} \dd\tau \, \tau^{r-1} {\rm Tr} K_{\hat{O}(g)}(\tau)
\,. 
\end{align}
The heat-kernel is the solution of the diffusion equation $(\partial_\tau +\hat{O}(g)) K_{\hat{O}(g)}(\tau;x,x')=0$ with initial condition $K_{\hat{O}(g)}(0;x,x')=\delta^{(d)}(x,x')$.
If the original classical action is Weyl invariant, we have that the operator $\hat{O}(g)^{-r}$ must transform with a bi-weight under Weyl \cite{Erdmenger:1997gy}, i.e., as
\begin{align}
\hat{O}(g+\delta g)
=
e^{(w-2)\sigma} \, {\hat{O}(g)} \, e^{-w\sigma} \,, 
\end{align}
where $w$ denotes the natural Weyl weight of the tensor field  $\delta \phi^i=w\sigma\phi^i$. For the quadratic action to be Weyl-invariant, we must have $-2w+2=d$ if we are working with second-order differential operators.\footnote{%
If the operator is rank $2n$, the instances of $2$ are replaced by $2n$, leading to a slightly different coefficient for the anomaly. We choose to give all formulas for the second-order case, but the generalization is straightforward assuming that a local expansion holds, as we also mention in due time. The local expansion of higher derivative operators is always assumed to work in the coincident limit $x=x'$, but the complications observed in Ref.~\cite{Barvinsky:2021ijq} at the level of bilocal tensors suggest that the problem should be treated with more care.
}

We can express a perturbation with respect to the metric, $\zeta_{\hat{O}} (g+\delta g;r)$, as
\begin{align}
\zeta_{\hat{O}} (g+\delta g;r)
&=
{\rm Tr} \big\{ {\hat{O}(g+\delta g)^{-r}} \big\}
=
{\rm Tr}  \big\{ \big[ ( 1 + \sigma )^{w-2} \,  {\hat{O}}(g) \, (  1 + \sigma  )^{-w}   + O(\sigma^2)   \big]^{-r}  \big\}\\ \nonumber
&=
{\rm Tr}  \bigg\{  {\hat{O}}(g)^{-r} + r  \,  { {\hat{O}}(g)^{-(s+1)}}     \bigg( (2-w) \sigma\,  {\hat{O}(g) }   + w  \,{\hat{O}}(g)  \sigma \bigg)       + O(\sigma^2)            \bigg\}\\ \nonumber
&=
 {\zeta_{\hat{O}}(g;r)}+2r\, {\rm Tr}  \big\{    \sigma \,   {\hat{O}}(g) ^{-r} + O(\sigma^2) \big\} \,.
\end{align}
Consequently, the variation of the effective action under the Weyl transformation is
\begin{align}
\delta_\sigma \Gamma[g]
=
\Gamma[g+\delta_\sigma g]
-
\Gamma[g]
=
{\rm Tr}  \big\{ \sigma \,{\hat{O}}(g)^{-r} \big\} |_{r=0}
\,,   
\end{align}
which can be evaluated as
\begin{align}
{\rm Tr}  \big\{  \sigma \,   {\hat{O}}(g) ^{-r}  \big\} |_{r=0}
=
 \int \dd^d x  \sqrt{g} \, \sigma(x)\, \langle x | \hat{O}(g)^{-r} | x \rangle   |_{r=0} 
=
 \int \dd^d x \sqrt{g} \, \sigma(x)  \, \zeta (g;0|x,x) \,. 
\end{align}
Using the relationship between the zeta function and the heat kernel, along with the \emph{asymptotic} small-$\tau$ Seeley-DeWitt expansion in the coincident limit, which tells us that $K_{\hat{O}(g)}(\tau;x,x) \sim \frac{1}{(4\pi \tau)^{d/2}}\sum_{n\geq0} a_n(x) \tau^n$ \cite{Vassilevich:2003xt}, one finds
\begin{align}
\zeta (g;0|x,x)
=
\frac{1}{(4\pi)^{d/2}} \, a_{d/2}(x) \,.
\end{align}
The convenience in using the heat kernel expansion is that the coefficients $a_{n}(x)$ can be computed algorithmically.
This suggests that the expression could be generalized to non-minimal
and higher-rank operators by employing the corresponding Seeley-DeWitt coefficients, and the only requirement is that an asymptotic expansion like Seeley-DeWitt's holds in the limit $x=x'$. This is known to hold, for example,
for higher derivative minimal operators \cite{Barvinsky:2019spa},
but much less is known for nonminimal operators (for example, those that appear in Refs.~\cite{Erdmenger:1997wy,Paci:2023twc} that are necessary for various tensor fields) even though it is reasonable to assume to still hold.

To arrive at our original objective, we compare the two variations with respect to $\sigma$, i.e., the one which made Duff feel ``uneasy'' in Ref.~\cite{Duff:1977ay}
\begin{align}
\delta_\sigma \Gamma
=
\int \dd^d x  \sqrt{g} \,
 \sigma(x) \, \langle T  \rangle\,,
\end{align}
with the one derived using an asymptotic series
\begin{align}
\delta_\sigma \Gamma
=
\frac{1}{(4\pi)^{d/2}}  \int \dd^d x  \sqrt{g}  \, \sigma(x) \, a_{d/2}(x)\,,
\end{align}
from which we obtain the integrated anomaly. Up to boundary terms, that are associated with the scheme-dependent $a'$ anomalies anyway, the nonintegrated anomaly is given by
\begin{align}
\langle T  (x)\rangle
=
\frac{1}{(4\pi)^{d/2}} \, a_{d/2}(x)\,,
\end{align}
which provides a mean to explicitly compute any anomaly parametrized in the main text.


\end{document}